\documentclass[a4paper,12pt]{article}
\usepackage{amssymb,amsmath,amsthm,mathrsfs,mathtools}
\usepackage{caption}
\usepackage{cite}
\usepackage[colorlinks]{hyperref}
\usepackage{eqparbox,graphicx}
\usepackage[affil-it]{authblk}
\usepackage{pgf,tikz,pgfplots,tkz-euclide,tikz-cd}
\usetikzlibrary{arrows,positioning,calc,fadings,decorations.pathreplacing,decorations,shapes.geometric,fit}
\tikzcdset{row sep/normal=7pt, column sep/normal=7pt}
\def\R{\mathbf{R}}
\newcommand\C{\mathbf{C}}
\let\kk\C
\tikzstyle{block} = [rectangle, draw, fill=blue!80!black!20, 
    text width=6.3em, text centered, rounded corners, minimum height=3em]
\newcommand\nodeone[1]{#1}
\newcommand\nodetwo[2]{\begin{smallmatrix}#1\\\scriptstyle #2\end{smallmatrix}}
\newcommand\nodethree[3]{\begin{smallmatrix}
\scriptstyle #1\\\scriptstyle #2\\\scriptstyle #3
\end{smallmatrix}}
\newcommand\Z{\mathbf{Z}}

\newcommand\rep[1]{{\operatorname{rep} #1}}
\newcommand\db[1]{\mathrm D^{\mathrm b}(\operatorname{rep} #1)}
\newcommand\zz[1]{\Z{Q}}
\newcommand\Hom[2][\empty]{\operatorname{Hom}_{#1}\left({#2}\right)}

\def\A{\mathcal{A}}

\newcommand\dg[1]{\mathcal{D}^{\geq #1}}
\newcommand\dl[1]{\mathcal{D}^{\leq #1}}

\newcommand\eq[1]{(\ref{#1})}

\def\id{\mathrm{id}}

\newcommand\rt{\longrightarrow}

\def\pp{{\mathbf{p}}}
\def\xx{{\mathbf{x}}}

\theoremstyle{definition}

\newcommand\heart[1]{\langle #1\rangle}

\newcommand\cq[1]{{\mathcal{C}}_{{#1}}}
\newcommand\dq[1]{{\mathcal{D}}_{{#1}}}

\newdimen\polrad
\title{Learning scattering amplitudes by heart}
\author{Severin Barmeier\thanks{email: severin.barmeier@math.uni-freiburg.de}~} 
\affil{Hausdorff Research Institute for Mathematics, Poppelsdorfer Allee 45,
53115 Bonn, Germany \authorcr \it Albert-Ludwigs-Universität Freiburg, Ernst-Zermelo-Str.~1, 79104 Freiburg im Breisgau, Germany}
\author{Koushik Ray\thanks{email: koushik@iacs.res.in}}
\affil{Indian Association for the Cultivation of Science,\authorcr
\it Calcutta 700 032, India}
\date{}
\begin{document}
\maketitle
\begin{abstract}
\noindent
The canonical forms associated to scattering amplitudes of planar 
Feynman diagrams
are interpreted in terms of masses of projectives, defined as the modulus of
their central charges, in the hearts of certain 
$t$-structures of derived categories of quiver representations and, 
equivalently, in terms of cluster tilting objects of the corresponding 
cluster categories. 
\end{abstract}
\setcounter{page}{0}
\thispagestyle{empty}
\clearpage
\section{Introduction}
The amplituhedron program of $\mathcal N = 4$ supersymmetric 
Yang--Mills theories \cite{aht} culminating in the ABHY construction 
\cite{abhy} has provided renewed 
impetus to the study of computation of scattering amplitudes in quantum field
theories, even without supersymmetry, using geometric ideas. 
While homological methods for the
evaluation of Feynman diagrams have been pursued for a long time \cite{hht},
the ABHY program associates the geometry of Grassmannians to the
amplitudes \cite{hbcgpt}. Relations of amplitudes to a variety of
mathematical notions and structures have been unearthed \cite{dfgk,miz,blr}. 
For scalar field theories the amplitudes are expressed 
in terms of Lorentz-invariant Mandelstam variables. The
space of Mandelstam variables is known as the kinematic space. The
combinatorial structure of the amplitudes associated with the planar Feynman
diagrams of the cubic scalar field theory
is captured by writing those as
differential forms associated to a polytope in the kinematic space, called the
associahedron, and their integrals. The form is named canonical form. 

\begin{figure}[h]
\centering{
\begin{tikzpicture}[x=12pt, y=12pt]
\draw (0,0) -- (0,-1); 
\begin{scope}[rotate around={45:(0,0)}]
\draw (0,0) -- (0,3);
\draw (0,0) -- (3,0);
\draw (1,0) -- (1,2); 
\draw (2,0) -- (2,1);
\end{scope}
\begin{scope}[xshift=5.8em,yshift=.5em]
\node at (-3,0) {$=$};
\draw(0,0) -- (0,1); 
\draw(0,0) -- (1,0); 
\draw(0,0) -- (-1,0); 
\draw(-1,0) -- (-2,1);
\draw(-1,0) -- (-2,-1);
\draw(1,0) -- (2,1);
\draw(1,0) -- (2,-1);
\path[<->] (3,0) edge node[above=-.2ex,font=\scriptsize] {dual} (5.5,0);
\end{scope}
\begin{scope}[xshift=14.8em,yshift=.5em]
\draw[black!30!white] (0:0) -- ++(162:.9) -- ++(126:1.2);
\draw[black!30!white] (0:0) -- ++(18:.9) -- ++(54:1.2);
\draw[black!30!white] (0:0) ++(162:.9) -- ++(198:1.2);
\draw[black!30!white] (0:0) ++(18:.9) -- ++(-18:1.2);
\draw[black!30!white] (0:0) -- ++(-90:1.9);
\draw (18:1.8) -- (90:1.8) -- (162:1.8) -- (234:1.8) -- (306:1.8) -- cycle;
\draw (234:1.8) -- (90:1.8);
\draw (90:1.8) -- (306:1.8);
\end{scope}
\end{tikzpicture}
}
\caption{Binary tree, Feynman diagram and triangulation of polygon}
\label{tfp}
\end{figure}
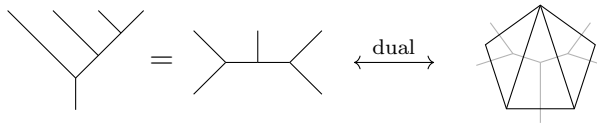

In an attempt to categorify the ideas we extract the canonical form
from cluster tilting objects in the cluster category of a quiver of type $A$ and, equivalently, from torsion pairs for the category of quiver representations and their intermediate $t$-structures.
Such a connection may be anticipated from the existing literature, 
but here we highlight the categorical framework available in the 
representation-theoretic literature, which we
expect to prove instructive in generalizing the computation of 
scattering amplitudes in general. We
restrict our attention to the planar tree level Feynman diagrams in a
cubic scalar field theory. These diagrams can be thought of as rooted
binary trees. The Feynman diagrams are dual
to triangulated polygons in the sense that they are obtained 
by drawing lines intersecting the
edges of a triangulated polygon as indicated in Fig.~\ref{tfp}.

\begin{figure}[h]
\begin{center}
\begin{tikzpicture}[node distance = 1.2in, auto]
\node[block] (feyn) {Feynman diagram};
\node[block] (tree) at ([yshift=-.8in]feyn) {Tree};
\node[block] (quiv) at ([xshift=3.2in]feyn) {Quiver $Q$};
\node[block] (clus) at ([xshift=.8in,yshift=-.8in]quiv) {Cluster category $\cq{Q}$};
\node[block] (tilt) at ([yshift=-.8in]clus) {Cluster tilting objects};
\node[block] (deri) at ([xshift=-.8in,yshift=-.8in]quiv) {Derived category $\dq{Q}$};
\node[block] (hear) at ([yshift=-.8in]deri) {Intermediate $t$-structures};
\node[block] (asso) at ([xshift=1.6in,yshift=-2.4in]feyn) {Associahedron};
\node[block] (form) at ([xshift=1.6in]asso) {Canonical form};
\node[block] (poly) at ([xshift=1.6in]feyn) {Triangulation};
\draw (tree) -- (feyn);
\draw (feyn) -- (poly);
\draw (poly) -- (asso);
\draw (poly) -- (quiv);
\draw (quiv) -- (clus);
\draw (quiv) -- (deri);
\draw (deri) -- (clus);
\draw (asso) -- (form);
\draw (hear) -- (tilt);
\draw (clus) -- (tilt);
\draw (deri) -- (hear);
\draw[->] (deri) -- (clus);
\draw[->,red] (tilt) -- (form);
\draw[->,red] (hear) -- (form);
\end{tikzpicture}
\end{center}
\caption{Flow chart}
\label{flowchart}
\end{figure}
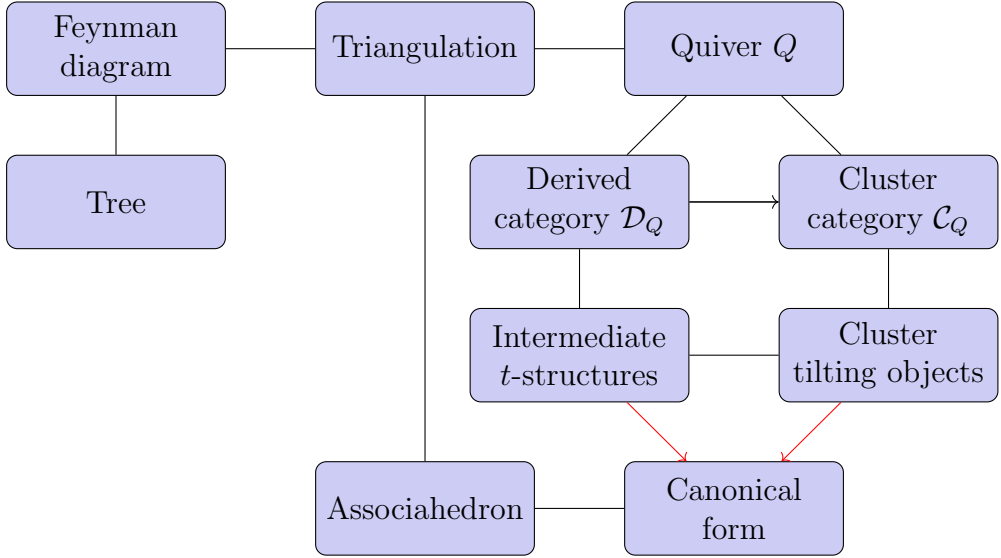

The various triangulations of a plane polygon, and hence the Feynman
diagrams, are in one-to-one correspondence with the vertices of
an associahedron, which also appears
in the representation theory of quivers \cite{schiffler}
in the form of an exchange graph \cite{by,qiu12,kq}. 
The canonical form is identified as the volume of the polytope dual to the
associahedron \cite{abhy,dfgk}. 
The vertices of the associahedron are also
associated to the set of cluster variables of cluster algebras \cite{ch}, 
two adjacent vertices being related
by a mutation \cite{keller}. In a categorical framework cluster algebras
are associated to cluster categories which 
can be realized as triangulated
orbit categories of the bounded derived categories of
quiver representations. 
We show that the canonical form is directly obtained from the central
charges of the projective indecomposable objects of hearts of intermediate $t$-structures. In other words, we obtain the red arrows in Fig.~\ref{flowchart}.

Let us describe the proposed interpretation at the outset. We shall then work
out two examples, which are easily generalized. In a perturbative
treatment of the
scattering of $N$ particles in quantum field theory, the conservation of
total momenta carried by the particles as well as the nature of interactions
are encoded in Feynman diagrams.
The contribution to the scattering amplitude corresponding to the diagrams
are expressed in terms of Lorentz-invariants formed out of the momenta of the
particles. Consideration of the diagrams beyond tree level, 
which involve integration over momenta, will be postponed to future work.
A set of such invariants, called planar variables, are
denoted $X_{ij}$, with $i$ and $j$ running over the
labels of $N$ particles. The indices are defined modulo $N$.
This, in addition to the symmetry
of the planar variables under the exchange of the indices, gives 
the set of planar variables a periodic structure. 
A certain combination of the planar variables, namely, the discrete Laplacian
operating on $X_{ij}$, relates to the Mandelstam variables.
This combination may be presented as a diamond-like relation with 
arrows indicating the sign of terms in the discrete Laplacian.
Considered seriatim for all the particles, the diamond 
can be knit into a mesh. 
If the mesh is continued ad infinitum it yields the Auslander--Reiten
(AR) quiver of the bounded derived category $\dq{Q} = \db{Q}$ of finite-dimensional
representations of a quiver $Q$ of type $A$ \cite{causal,bdmty}, the planar
variables corresponding one-to-one with indecomposable objects and the arrows in the 
mesh taken to represent the irreducible morphisms in $\dq{Q}$.
The periodicity of the planar variables further restricts the structure by
dictating an identification of objects in the AR quiver. 
We observe that this identification happens to correspond precisely to the passage from the derived category $\dq{Q}$ to the cluster category $\cq{Q}$, which is the triangulated category of orbits under a certain autoequivalence \cite{keller}.

Planar tree level Feynman diagrams of 
$N$ particles are dual to triangulations of an $N$-gon, which in turn can be related to the cluster algebra or the cluster category of the $A_{N-3}$ quiver \cite{fominzelevinsky,ccs}, the combinatorial structure of the set of all triangulations being described by mutations in the cluster algebra or cluster category. In the categorical setting, these combinatorics can be expressed via a range of different structures. In this article we focus on two such structures: intermediate $t$-structures for the derived category $\dq{Q}$ and cluster tilting objects for the cluster category $\cq{Q}$. Our goal is to use the correspondences
\[
\text{Feynman diagrams} \longleftrightarrow \text{intermediate $t$-structures in } \dq{Q} \longleftrightarrow \text{cluster tilting objects in } \cq{Q}
\]
to obtain the explicit expression of the canonical form.

The similarity between the mesh and the AR quiver is formal
so far. While the objects in the cluster category are orbits of complexes of
quiver representations in the derived category,
the planar variables are real numbers. In order to relate these we remark that
the central charge of objects are complex numbers
associated to equivalence classes of the objects in the Grothendieck group of
the derived category. These are related through the mesh relations, which coincide with the mesh relations among the planar
variables $X_{ij}$. By identifying the modulus of the 
central charge of an object in the AR quiver of the cluster category with
the corresponding planar variable $X_{ij}$ we re-derive, not surprisingly, the relations among the latter from 
the mesh relations of the derived category. The modulus of central charge is
called the mass \cite{bridgeland}. In here it relates to the squared
invariant mass of a collection of particles. In the categorical parlance
$X_{ij}$ are interpreted as 
the masses of indecomposable objects in the derived category. 
Their relations as derived from momentum
conservations descend from the central charges on the stability lines. 
Each cluster tilting object in the cluster category
corresponds to a Feynman diagram. The contribution of each Feynman diagram to
the amplitude is thus given by a term in the canonical form.
Each term is a 
logarithmic $(N-3)$-form written in terms of the planar variables. 
We show that from the point of view of derived categories each such term is given by the masses of projective objects 
of the hearts of intermediate $t$-structures.

Categorical formulation facilitates organizing computations.
The present formulation, while not as geometric as the associahedron or the 
Grassmannians, has the virtue of being algebraic and seems to be of use in
developing computer algorithms for the evaluation of the tree diagrams.
We now exemplify the program chalked out above for two examples, $N=5$ and
$N=6$, the latter being completely generic. In the next section we review the
derivation of the mesh relations among the planar variables from momentum
conservation.  The case of $N=5$ particles is worked out explicitly. 
In section \ref{sec:dercat} we recall the notions of derived categories,
intermediate $t$-structures and cluster categories and central charges,
exemplified for the case of $A_2$ quiver, corresponding to the case of $N=5$
particles. The canonical form for this case is then obtained for this case.
In section \ref{sec:6} we present the case of $N=6$ particle scattering
corresponding to the $A_3$ quiver. We obtain the mesh relations, the cluster
category and retrieve the canonical form, before concluding in section
\ref{sec:concl}.
The two examples are described in the following two sections.
In each case we first identify the cluster category from the mesh 
relations of the planar variables. We then obtain the cluster tilting objects and
identify their direct summands, which may be viewed equivalently as projective
objects of hearts of intermediate $t$-structures. Their masses are then shown to 
correspond to terms of the canonical form.
\section{Kinematics}
Let us recall the definition of kinematic variables
for  the scattering of a system of $N$ scalar particles \cite{abhy}. Their
 momenta are vectors in $\R^{1,3}$, 
denoted $\pp_i$, for $i=1,2,\dotsc,N$,
satisfying the conservation equation 
\begin{equation}
\label{mom:cons}
\sum_{i=1}^N\pp_i=0.
\end{equation} 
This is solved by writing the momenta in terms of another set of $N$
four-vectors $\xx$ as
\begin{equation}
\label{p2x}
\pp_i = \xx_{i+1}-\xx_i,
\end{equation} 
where from now on we define the indices modulo $N$, in particular,
$\xx_{N+1}=\xx_1$.
Mandelstam variables are quadratic invariants for a pair of particles
\begin{equation}
\label{def:s}
s_{ij}=(\pp_i+\pp_j)^2
\end{equation} 
where the norm of a four-vector $\pp = (p_0, p_1, p_2, p_3)$ is
defined as
$\pp^2=-p_0^2+p_1^2+p_2^2+p_3^2$.
Entities defined similarly with the $\mathbf x$'s as 
\begin{equation}
\label{x2X}
X_{ij}=(\xx_i-\xx_j)^2
\end{equation} 
are called planar variables. 
These are symmetric with respect to exchange of indices by
definition, $X_{ji}=X_{ij}$. Using  \eq{p2x} and the
periodicity of the indices the planar variables are related to
the Mandelstam variables as
\begin{equation} 
\label{X2s}
s_{ij}=\pp_i^2+\pp_j^2+X_{i,j+1}+X_{i+1,j}-X_{i,j}-X_{i+1,j+1}.
\end{equation} 
If we now assume that the particles are massless, that is, the momentum vectors
are null, $\pp_i^2=0$ for each
$i$, then
\begin{equation}
\label{sX:null}
s_{ii}=2\pp_i^2=0,\qquad X_{i,i+1}=\pp_i^2=0
\end{equation} 
and the relation \eq{X2s} becomes
\begin{equation}
\label{sX}
s_{ij}=X_{i,j+1}+X_{i+1,j}-X_{ij}-X_{i+1,j+1}.
\end{equation} 
The right hand side is the negative discrete Laplacian operating on the
planar variables.  In particular, we have
\begin{equation}
\label{splx}
\begin{aligned}
X_{N,N+1} &= \eqmakebox[xN1]{$X_{N,1}$} = \eqmakebox[xN1]{$X_{1,N}$} = 0, \\
X_{2,N+1} &= \eqmakebox[xN1]{$X_{2,1}$} = \eqmakebox[xN1]{$X_{1,2}$} = 0.
\end{aligned}
\end{equation}
Equation \eq{sX} can be pictorially presented as 
\begin{equation}
\label{mesh:X}
s_{ij}=\begin{tikzpicture}[baseline=-2.6pt]
\matrix (m) [matrix of math nodes, row sep={3em,between origins}, text height=1.4ex, 
column sep={3em,between origins}, text depth=.15ex, ampersand replacement=\&] 
{ \& X_{i,j+1} \& \\
X_{ij} \&\& X_{i+1,j+1} \\
\& X_{i+1,j} \&\\
};
\draw[->] (m-2-1) -- (m-1-2);
\draw[->] (m-2-1) -- (m-3-2);
\draw[->] (m-1-2) -- (m-2-3);
\draw[->] (m-3-2) -- (m-2-3);
\end{tikzpicture}
\end{equation} 
For any given value of $N$ this unit can be used to weave a mesh
\cite{causal}, which, upon using the periodicity of the indices, the symmetry
of $X$'s and equation \eq{splx}, gives rise to the cluster category of the $A_{N-3}$
quiver. In the next two sections we work out the examples of $N=5$ and $N=6$
and obtain the canonical forms from the central charges.
\subsection{Example of $N=5$}
\label{sec:N5}
In this section we recall various notions pertaining to the categorical
description of the scattering amplitude. In order to be explicit we shall
often use the example of an $A_2$ quiver arising in the case of scattering of
five particles. 
The mesh diagram knit from \eq{mesh:X} is 
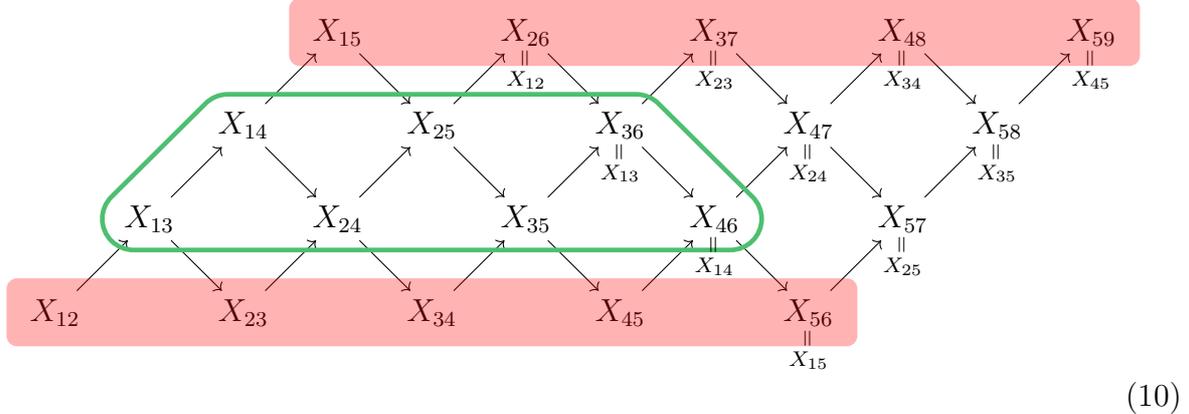
\begin{equation} 
\label{mesh5:1}
\begin{tikzpicture}[baseline=-4pt]
\matrix (m) [matrix of math nodes, row sep={3em,between origins}, text height=1.4ex, 
column sep={3em,between origins}, text depth=.15ex, ampersand replacement=\&] 
{
\&\&\& X_{15} \&\& X_{26} \&\& X_{37} \&\& X_{48} \&\& X_{59}  \\
\&\& X_{14} \&\& X_{25} \&\& X_{36} \&\& X_{47} \&\& X_{58} \& \\
\& X_{13} \&\& X_{24} \&\& X_{35} \&\& X_{46} \&\& X_{57} \&\& \\
 X_{12} \&\& X_{23} \&\& X_{34} \&\& X_{45} \&\& X_{56} \&\&\& \\
};
\foreach \i/\j in {1/2,3/4,5/6,7/8,9/10}
\draw[->] (m-4-\i)--(m-3-\j);
\foreach \i/\j in {2/3,4/5,6/7,8/9,10/11}
\draw[->] (m-3-\i)--(m-2-\j);
\foreach \i/\j in {3/4,5/6,7/8,9/10,11/12}
\draw[->] (m-2-\i)--(m-1-\j);
%
\foreach \i/\j in {4/5,6/7,8/9,10/11}
\draw[->] (m-1-\i)--(m-2-\j);
\foreach \i/\j in {3/4,5/6,7/8,9/10}
\draw[->] (m-2-\i)--(m-3-\j);
\foreach \i/\j in {2/3,4/5,6/7,8/9}
\draw[->] (m-3-\i)--(m-4-\j);
%
\begin{scope}[xshift=-13.5em,yshift=-1.5em]
\draw[line width=.15em,black!20!green!80!blue!70,line cap=round] (0,-1em) -- ++(18.5em,0) arc[start angle=-90, end angle=45, radius=1em] -- ++(-3em,3em) arc[start angle=45, end angle=90, radius=1em] -- ++(-13em,0) arc[start angle=90, end angle=135, radius=1em] -- ++(-3em,-3em) arc[start angle=135, end angle=270, radius=1em] -- cycle;
\end{scope}
\node[font=\scriptsize,below=-1.3ex of m-1-6]  {$\overset{\rotatebox{90}{=}}{X_{12}}$};
\node[font=\scriptsize,below=-1.3ex of m-1-8]  {$\overset{\rotatebox{90}{=}}{X_{23}}$};
\node[font=\scriptsize,below=-1.3ex of m-1-10] {$\overset{\rotatebox{90}{=}}{X_{34}}$};
\node[font=\scriptsize,below=-1.3ex of m-1-12] {$\overset{\rotatebox{90}{=}}{X_{45}}$};
\node[font=\scriptsize,below=-1.3ex of m-2-7]  {$\overset{\rotatebox{90}{=}}{X_{13}}$};
\node[font=\scriptsize,below=-1.3ex of m-2-9]  {$\overset{\rotatebox{90}{=}}{X_{24}}$};
\node[font=\scriptsize,below=-1.3ex of m-2-11] {$\overset{\rotatebox{90}{=}}{X_{35}}$};
\node[font=\scriptsize,below=-1.3ex of m-3-8]  {$\overset{\rotatebox{90}{=}}{X_{14}}$};
\node[font=\scriptsize,below=-1.3ex of m-3-10] {$\overset{\rotatebox{90}{=}}{X_{25}}$};
\node[font=\scriptsize,below=-1.3ex of m-4-9]  {$\overset{\rotatebox{90}{=}}{X_{15}}$};
\node[fill=red, rounded corners, fit=(m-4-1.north west) (m-4-9.south east),
inner sep=.4em,opacity=.3] (dleq) {};
\node[fill=red, rounded corners, fit=(m-1-4.north west) (m-1-12.south east),
inner sep=.4em,opacity=.3] (dleq) {};
\end{tikzpicture}
\end{equation} 
Using the periodicity of the indices modulo $5$ and the symmetry
$X_{ji}=X_{ij}$, for example $X_{59}=X_{54}=X_{45}$, 
the ones in red blocks are null by \eq{sX:null} and \eq{splx}. 
The canonical form associated to the $N=5$ amplitude is  \cite{abhy}
\begin{multline}
\label{can:5}
\Omega_5 
= d\log X_{14}\wedge d\log X_{13}
-d\log X_{35}\wedge d\log X_{13}
+d\log X_{35}\wedge d\log X_{25}\\
-d\log X_{24}\wedge d\log X_{25}
+d\log X_{24}\wedge d\log X_{14}.
\end{multline}
The scattering amplitude is obtained from the canonical form using relations
among the planar variables \cite{abhy}. These relations are derived from the
mesh relations in the derived category. 
Let us indicate the combinatorial scheme for fixing the relative signs 
of the terms. The set of planar variables appearing in each term of the 
canonical form are identified first and one term is fixed, say, 
the first one in \eq{can:5} with a positive sign.
We then replace one of the planar variables with a new one. The new term is
given a negative sign, as in the second term in \eq{can:5}, where $X_{14}$ of
the first term is replaced with $X_{35}$. Repeating this we obtain the full
canonical form. 

The main observation described in this article is that 
after dispensing with the red blocks along with the arrows to and from the
corresponding nodes we are left with the Auslander--Reiten (AR) quiver of the bounded derived category 
of the finite-dimensional representations of the $A_2$ quiver. Moreover, the green block in
\eq{mesh5:1} is, after identifying the objects as indicated, precisely the cluster category of $A_2$, and the terms of the
canonical form are in one-to-one correspondence with the cluster tilting objects and, on the level of the bounded derived category, with the intermediate $t$-structures.
The next three subsections digress to recall various notions 
pertaining to the categorification of the canonical form. The discussions are
brief and often through examples. We refer to \cite{thomas,keller,subir} 
for further elaborations. 
\section{Categorical picture}
\label{sec:dercat}
A quiver is a directed graph, that is, a collection of vertices and arrows amongst them. A representation of
a quiver is given by associating a finite-dimensional vector space to the vertices and
linear maps between these vector spaces to the arrows. An $A_n$ quiver, is a quiver whose underlying undirected graph is a Dynkin diagram of type $A_n$. In this article we work with the $A_n$ quiver with linear orientation
\begin{equation}
1\rt 2\rt\cdots\rt n,
\end{equation} 
the natural numbers labelling the vertices. Such representations form an Abelian category and every representation is isomorphic to a direct sum of indecomposable representations. For a quiver of type ADE, Gabriel's Theorem states that there are only finitely many indecomposable representations, which for the linearly oriented $A_n$ quiver are of the simple form
\[
0 \rt \cdots \rt 0 \rt \C \stackrel{\id}{\rt} \cdots \stackrel{\id}\rt \C \rt 0 \rt \cdots \rt 0
\]
where the vector space at each node is taken to be either
the $1$-dimensional vector space of complex numbers $\C$, or the trivial vector space $0$ and the linear maps between copies of $\C$ are the identity maps.

For example, the $A_2$ quiver is $1 \rt 2$. The Abelian category of 
finite-dimensional representations $\rep{A_2}$
contains three indecomposable representations
\begin{equation}
\kk\rt 0, \qquad 0\rt\kk \qquad\text{and}\qquad \kk\stackrel{\id}\rt\kk.
\end{equation}
These representations are also denoted $1$, $2$ and $\nodetwo{1}{2}$, 
respectively. Here a single number $i$ denotes the simple representation 
with a one-dimensional vector space at the vertex labelled $i$ in the quiver 
and $\nodetwo{i}{j}$ denotes an extension of $i$ by $j$.
These three representations fit into a short exact sequence
\begin{equation}
\label{SES:2}
0\rt \nodeone{2}\rt \nodetwo{1}{2}\rt \nodeone{1}\rt 0
\end{equation}
in $\rep{A_2}$. 

A derived category is obtained from an Abelian category by promoting short
exact sequences to triangles. Elements of the derived category are $\mathbb Z$-graded cochain complexes of elements of the Abelian category, where two complexes are considered isomorphic in the derived category 
if there is a chain of morphisms of complexes between the two inducing an isomorphism in
cohomology. The bounded derived category is formed by considering only complexes which are zero in all but finitely many degrees. The
derived category is equipped with a \emph{shift functor} $[1]$ which, when applied
to a complex, shifts all the elements of the complex down by one degree. If $C^i$ denotes the degree $i$ component of a complex $C^{\bullet}$, then the degree $i$th chain in the shifted complex is 
$C^{\bullet}[n]^i=C^{i+n}$, where $[n]$ is the $n$-fold composition of $[1]$.

Given an element $M$ in the Abelian category, one can view it as a complex with $M$ placed in degree $0$ and 
$M[n]$ denotes its shift by $n$, e.g.\
\begin{equation}
\label{shift}
\begin{tikzpicture}[baseline=-4pt]
\matrix (m) [matrix of math nodes, row sep=.25em, text height=1.4ex, column sep={3em,between origins}, text depth=.15ex, ampersand replacement=\&, row 1/.style={font=\scriptsize}]
{
\&[1em] \& \text{\scriptsize $-2$} \& {\scriptsize -1} \& {\scriptsize 0} \& {\scriptsize 1} \& \\
\eqmakebox[M1]{$M$}    \&[1em] (\dotsb \& 0 \& 0
\& M \& 0 \& \dotsb) \\
\eqmakebox[M1]{$M[1]$} \&[1em] (\dotsb \& 0 \& M
\& 0 \& 0 \& \dotsb)\rlap{.} \\
};
\draw[<->] (m-2-1) -- (m-2-2);
\draw[<->] (m-3-1) -- (m-3-2);
\draw[->] (m-2-2) -- (m-2-3);

\draw[->] (m-2-3) -- (m-2-4);
\draw[->] (m-2-4) -- (m-2-5);
\draw[->] (m-2-5) -- (m-2-6);
\draw[->] (m-2-6) -- (m-2-7);
\draw[->] (m-3-2) -- (m-3-3);
\draw[->] (m-3-3) -- (m-3-4);
\draw[->] (m-3-4) -- (m-3-5);
\draw[->] (m-3-5) -- (m-3-6);
\draw[->] (m-3-6) -- (m-3-7);
\end{tikzpicture}
\end{equation}

Given a short exact sequence $0\rt A\stackrel{f}\rt B\stackrel{g}\rt C\rt 0$ in an Abelian category,
the two complexes 
\begin{equation}
\begin{tikzpicture}[baseline=-4pt]
\matrix (m) [matrix of math nodes, row sep=.25em, text height=1.4ex, column sep={3em,between origins}, text depth=.15ex, ampersand replacement=\&, row 1/.style={font=\scriptsize}]
{
                                \& -2 \& -1 \& 0 \& 1 \& \\
\dotsb                          \& 0  \& A  \& B \& 0 \& \dotsb \\
\llap{and \hspace{1.5em}}\dotsb \& 0  \& 0  \& C \& 0 \& \dotsb \\
};
\draw[->] (m-2-1) -- (m-2-2);
\draw[->] (m-2-2) -- (m-2-3);
\draw[->] (m-2-3) -- (m-2-4);
\draw[->] (m-2-4) -- (m-2-5);
\draw[->] (m-2-5) -- (m-2-6);
\draw[->] (m-3-1) -- (m-3-2);
\draw[->] (m-3-2) -- (m-3-3);
\draw[->] (m-3-3) -- (m-3-4);
\draw[->] (m-3-4) -- (m-3-5);
\draw[->] (m-3-5) -- (m-3-6);
\end{tikzpicture}
\end{equation}
are isomorphic in the derived category, as the maps in the short exact sequence give a map of complexes which induces an isomorphism in cohomology. Hence, there is a map $C \rt A[1]$ in the derived category, given by
\begin{equation}
\begin{tikzpicture}[baseline=-4pt]
\matrix (m) [matrix of math nodes, row sep={3em,between origins}, text height=1.4ex, column sep={3em,between origins}, text depth=.15ex, ampersand replacement=\&]
{
\dotsb \& 0 \& A \& B \& 0 \& \dotsb \\
\dotsb \& 0 \& A \& 0 \& 0 \& \dotsb \\
};
\draw[->] (m-1-1) -- (m-1-2);
\draw[->] (m-1-2) -- (m-1-3);
\draw[->] (m-1-3) -- (m-1-4);
\draw[->] (m-1-4) -- (m-1-5);
\draw[->] (m-1-5) -- (m-1-6);
\draw[->] (m-2-1) -- (m-2-2);
\draw[->] (m-2-2) -- (m-2-3);
\draw[->] (m-2-3) -- (m-2-4);
\draw[->] (m-2-4) -- (m-2-5);
\draw[->] (m-2-5) -- (m-2-6);
\draw[->] (m-1-2) -- (m-2-2);
\draw[double equal sign distance,line width=.4pt] (m-1-3) -- (m-2-3);
\draw[->] (m-1-4) -- (m-2-4);
\draw[->] (m-1-5) -- (m-2-5);
\end{tikzpicture}
\end{equation}
where the lower complex is $A[1]$. This is expressed by drawing a triangle
\begin{equation}
\begin{tikzcd}
&B\arrow[dr]&\\[1.25em]
A\arrow[ur]&&C\arrow{ll}{[1]}
\end{tikzcd}
\end{equation} 
and such triangles equip the derived category with its triangulated structure. 
Another way to encode the information of the triangle is writing
the complex
\begin{equation}
\label{les}
\cdots \rt A \rt B\rt C\rt A[1]\rt B[1]\rt \cdots
\end{equation} 
in the derived category corresponding to the short exact sequence in the
Abelian category. The complex continues on both sides. 
In this fashion the Abelian category $\rep{A_n}$ gives rise to 
the bounded derived category, denoted $\dq{A_n} = \db{A_n}$.

Since the Abelian category $\rep{A_n}$ is hereditary (i.e.\ $\operatorname{Ext}^k_{\rep{A_n}}$ vanishes for $k \geq 2$) and only has only finitely many indecomposable objects, any object in the bounded derived category can be decomposed into complexes of the form $M [n]$ for some indecomposable representation $M$ of $\rep{A_n}$ and some integer $n$. All the essential information of the derived category $\dq{A_n}$ is then captured diagrammatically by its AR quiver, which encodes the morphisms and extensions between the indecomposable objects of the derived category. For example, starting from \eq{SES:2} we have, by \eq{les}, the long sequence
in $\dq{A_2}$, which is depicted as the AR quiver
\begin{equation}
\label{dbA}
\begin{tikzpicture}[baseline=-4pt]
\matrix (m) [matrix of math nodes, minimum width=1.4em, row sep={3.6em,between origins}, text height=1.4ex, column sep={3.6em,between origins}, text depth=.15ex, ampersand replacement=\&]
{
\& \nodetwo{1}{2} \&\& \nodeone{2}[1] \&\& \nodeone{1}[1] \&\&  \\
2 \&\& 1 \&\& \nodetwo{1}{2}[1] \&\& \nodeone{2}[2] \& \phantom{\hspace{2.6em}} \\
};
\foreach \i/\j in {1/2,3/4,5/6}
\draw[->] (m-2-\i)--(m-1-\j);
%
\foreach \i/\j in {2/3,4/5,6/7}
\draw[->] (m-1-\i)--(m-2-\j);
%
\draw[->,dashed] (m-1-6) -- (m-1-4) node[above,midway,font=\scriptsize] {$\tau$};
\draw[->,dashed] (m-1-4) -- (m-1-2) node[above,midway,font=\scriptsize] {$\tau$};
\draw[->,dashed] (m-2-7) -- (m-2-5) node[above,midway,font=\scriptsize] {$\tau$};
\draw[->,dashed] (m-2-5) -- (m-2-3) node[above,midway,font=\scriptsize] {$\tau$};
\draw[->,dashed] (m-2-3) -- (m-2-1) node[above,midway,font=\scriptsize] {$\tau$};
\end{tikzpicture}
\end{equation}
The diagram
continues \emph{ad infinitum} on both sides and the dashed arrows $\tau$ are the AR translations, which encode the triangles of $\dq{A_2}$.
This diagram can be identified with the middle portion, between the omitted red blocks, of the mesh \eq{mesh5:1},
where the nodes in the AR quiver correspond to the planar variables $X_{ij}$ in \eq{mesh5:1}.
\subsection{Intermediate $t$-structures}
Given the derived category $\mathcal D = \mathrm D^{\mathrm b} (\A)$ of an Abelian category $\A$, one can recover $\A$ as the complexes concentrated in degree $0$. This process of obtaining an Abelian category from a derived category is generalized and axiomatized in the notion of a $t$-structure which provide a way of vivisecting the derived category, called truncations, realizing the Abelian
categories as the portion common to the truncated parts, called \emph{hearts}. The original Abelian
category is usually referred to as the \emph{standard heart} and the
corresponding $t$-structure as the standard $t$-structure.


A $t$-structure on $\mathcal D$ is a pair $(\dl{0},\dg{0})$ of strictly full subcategories,
satisfying 
\begin{enumerate}
\item $\dl{0}\subset\dl{1}$ and $\dg{1}\subset\dg{0}$
\item $\operatorname{Hom}_{\mathcal D} (\dl{0}, \dg{1})=0$
\item for each object $M$ of the triangulated category $\mathcal D$, there exists a
distinguished triangle
\begin{equation}
\begin{tikzcd} 
& M_{\geq 1}\arrow[dashed,dr] & \\[1.25em]
M \ar[ur] & & M_{\leq 0}\ar[ll]
\end{tikzcd}
\end{equation} 
\end{enumerate}
where we set $\dg{m} = \dg{0}[-m]$ and $\dl{n} = \dl{n}[-n]$ for any integers $m$ and $n$. 
Further, a $t$-structure is called \emph{bounded} if
each $M$ in $\mathcal D$ is contained in $\dg{m}\cap\dl{n}$
for some integers $m$ and $n$. All $t$-structures in here are bounded.

When $\dl{0}$ is the full subcategory of complexes whose cohomology vanishes in positive degrees, and $\dg{0}$ the subcategory of complexes whose cohomology vanishes in negative degrees, then $(\dl{0},\dg{0})$ is called the standard $t$-structure and its heart $\dg{0} \cap \dl{0}$ is the standard heart and recovers the Abelian category $\A$.

So far, our discussion of $t$-structures have been rather general. 
We now briefly recall the notion of intermediate $t$-structures. A $t$-structure $(\mathcal D^{\preceq 0}, \mathcal D^{\succeq 0})$ is called intermediate with respect to the standard $t$-structure $(\dl{0}, \dg{0})$ if $\dl{0}[1] \subset \mathcal D^{\preceq 0} \subset \dl{0}$. These intermediate $t$-structures yield the terms of the
canonical form and can be obtained 
by tilting with respect to torsion pairs, which are pairs of subcategories for the initial Abelian category $\mathcal A$ \cite{by}. In order to be concrete we shall return to the specific examples of the $A_2$ and $A_3$ quivers. 

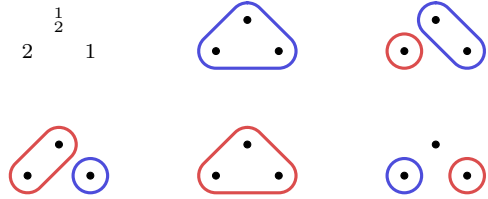
\begin{figure}
\begin{center}
\begin{tikzpicture}[x=1em,y=1em]
\node[font=\scriptsize] at (-1,0) {$\nodeone{2}$};
\node[font=\scriptsize] at  (0,1) {$\nodetwo{1}{2}$};
\node[font=\scriptsize] at  (1,0) {$\nodeone{1}$};
\begin{scope}[xshift=6em]
\draw[line width=.4pt,fill=black] (-1,0) circle(.25ex);
\draw[line width=.4pt,fill=black]  (0,1) circle(.25ex);
\draw[line width=.4pt,fill=black]  (1,0) circle(.25ex);
\draw[line width=.1em,blue!80!black!70,line cap=round] (0,1.55) arc[start angle=90, end angle=135, radius=.55em] -- ++(-1,-1) arc[start angle=135, end angle=270, radius=.55em] -- ++(2,0) arc[start angle=-90, end angle=45, radius=.55em] -- ++(-1,1) arc[start angle=45, end angle=90, radius=.55em];
\end{scope}
\begin{scope}[xshift=12em]
\draw[line width=.4pt,fill=black] (-1,0) circle(.25ex);
\draw[line width=.4pt,fill=black]  (0,1) circle(.25ex);
\draw[line width=.4pt,fill=black]  (1,0) circle(.25ex);
\draw[line width=.1em,red!80!black!70,line cap=round] (-1,0) circle(.55em);
\draw[line width=.1em,blue!80!black!70,line cap=round] (0,1.55) arc[start angle=90, end angle=225, radius=.55em] -- ++(1,-1) arc[start angle=-135, end angle=45, radius=.55em] -- ++(-1,1) arc[start angle=45, end angle=90, radius=.55em];
\end{scope}
\begin{scope}[yshift=-4em]
\draw[line width=.4pt,fill=black] (-1,0) circle(.25ex);
\draw[line width=.4pt,fill=black]  (0,1) circle(.25ex);
\draw[line width=.4pt,fill=black]  (1,0) circle(.25ex);
\draw[line width=.1em,blue!80!black!70,line cap=round] (1,0) circle(.55em);
\draw[line width=.1em,red!80!black!70, line cap=round] (0,1.55) arc[start angle=90, end angle=-45, radius=.55em] -- ++(-1,-1) arc[start angle=315, end angle=135, radius=.55em] -- ++(1,1) arc[start angle=135, end angle=90, radius=.55em];
\end{scope}
\begin{scope}[xshift=6em,yshift=-4em]
\draw[line width=.4pt,fill=black] (-1,0) circle(.25ex);
\draw[line width=.4pt,fill=black]  (0,1) circle(.25ex);
\draw[line width=.4pt,fill=black]  (1,0) circle(.25ex);
\draw[line width=.1em,red!80!black!70,line cap=round] (0,1.55) arc[start angle=90, end angle=135, radius=.55em] -- ++(-1,-1) arc[start angle=135, end angle=270, radius=.55em] -- ++(2,0) arc[start angle=-90, end angle=45, radius=.55em] -- ++(-1,1) arc[start angle=45, end angle=90, radius=.55em];
\end{scope}
\begin{scope}[xshift=12em,yshift=-4em]
\draw[line width=.4pt,fill=black] (-1,0) circle(.25ex);
\draw[line width=.4pt,fill=black]  (0,1) circle(.25ex);
\draw[line width=.4pt,fill=black]  (1,0) circle(.25ex);
\draw[line width=.1em,red!80!black!70,line cap=round] (1,0) circle(.55em);
\draw[line width=.1em,blue!80!black!70,line cap=round] (-1,0) circle(.55em);
\end{scope}
\end{tikzpicture}
\end{center}
\caption{The 5 torsion pairs $(\mathcal T, \mathcal F)$ of $\rep{A_2}$ with torsion class $\mathcal T$ (blue) and torsion-free class $\mathcal F$ (red)}
\label{torsionpairs2}
\end{figure}

Let $\dq{A_2} = \db{A_2}$ and let $(\dq{A_2}^{\leq 0}, \dq{A_2}^{\geq 0})$ denote the standard $t$-structure on $\dq{A_2}$ with heart $\dq{A_2}^{\leq 0} \cap \dq{A_2}^{\geq 0} \simeq \rep{A_2}$. A torsion pair $(\mathcal T, \mathcal F)$ for $\rep{A_2}$ may be obtained by choosing $\mathcal T$ to be the Abelian subcategory generated by any collection of indecomposable representations
which is closed under extensions and quotients, and then letting $\mathcal F =
\mathcal T^\perp = \{ Y \in \rep{A_2} \mid \Hom{X,Y} = 0, \ \forall X \in \mathcal T \}$ be its right orthogonal complement.
The Abelian category $\rep{A_2}$ has five torsion pairs which are presented
in Fig.~\ref{torsionpairs2} and the corresponding intermediate $t$-structures of $\dq{A_2}$ in Fig.~\ref{hearts2} where we omitted the arrows (cf.\ \eqref{abelianAR} and \eqref{dbA2}). The black dots correspond to the indecomposable objects of $\rep{A_2}$. In each diagram of Fig.~\ref{hearts2} the blue part corresponds to
$\dq{A_2}^{\preceq 0}$ and the red part to $\dq{A_2}^{\succeq 1}$ of the $t$-structure. The
vertices in the shaded part correspond to the indecomposable objects in the
heart of the $t$-structure, which is given by $\dq{A_2}^{\preceq 0} \cap
\dq{A_2}^{\succeq 0}$. Diagrammatically, the heart is obtained as the intersection of the blue part and the shift (given by glide reflection to the right) of the red part.

\begin{figure}[h]
\centering{
\begin{tikzpicture}[x=1.5em, y=1.5em]
\draw[line width=1ex,yellow!50!red!80!white!40, fill=yellow!50!red!80!white!40, line cap=round, rounded corners=.125ex] (-1,1) -- (0,2) -- (1,1) -- (0,0) -- cycle;
\draw[line width=.4pt] (-2,2) circle(.25ex);
\draw[line width=.4pt] (-2,0) circle(.25ex);
\draw[line width=.4pt,fill=black] (-1,1) circle(.25ex);
\draw[line width=.4pt]  (0,2) circle(.25ex);
\draw[line width=.4pt]  (0,0) circle(.25ex);
\draw[line width=.4pt]  (1,1) circle(.25ex);
\draw[line width=.4pt]  (2,0) circle(.25ex);
\draw[line width=.4pt]  (2,2) circle(.25ex);
\draw[line width=.4pt]  (3,1) circle(.25ex);
\draw[line width=.4pt]  (4,0) circle(.25ex);
\draw[line width=.4pt]  (4,2) circle(.25ex);
\node[font=\scriptsize] at (-1.4,1.03) {$P$};
\node[font=\scriptsize] at (1.05,1.9) {$\mathcal R (P)$};
\begin{scope}[shift={(13em,0)}]
\draw[line width=1ex,yellow!50!red!80!white!40, fill=yellow!50!red!80!white!40, line cap=round, rounded corners=.125ex] (0,0) -- (2,2);
\draw[line width=.4pt] (-2,2) circle(.25ex);
\draw[line width=.4pt] (-2,0) circle(.25ex);
\draw[line width=.4pt] (-1,1) circle(.25ex);
\draw[line width=.4pt]  (0,2) circle(.25ex);
\draw[line width=.4pt,fill=black]  (0,0) circle(.25ex);
\draw[line width=.4pt]  (1,1) circle(.25ex);
\draw[line width=.4pt]  (2,0) circle(.25ex);
\draw[line width=.4pt]  (2,2) circle(.25ex);
\draw[line width=.4pt]  (3,1) circle(.25ex);
\draw[line width=.4pt]  (4,0) circle(.25ex);
\draw[line width=.4pt]  (4,2) circle(.25ex);
\node[font=\scriptsize] at (-.4,.03) {$P$};
\node[font=\scriptsize] at (2.5,1.5) {$\mathcal R (P)$};
\end{scope}
\begin{scope}[shift={(26em,0)}]
\draw[line width=1ex,yellow!50!red!80!white!40, fill=yellow!50!red!80!white!40, line cap=round, rounded corners=.125ex] (0,2) -- (2,0);
\draw[line width=.4pt] (-2,2) circle(.25ex);
\draw[line width=.4pt] (-2,0) circle(.25ex);
\draw[line width=.4pt] (-1,1) circle(.25ex);
\draw[line width=.4pt,fill=black]  (0,2) circle(.25ex);
\draw[line width=.4pt]  (0,0) circle(.25ex);
\draw[line width=.4pt]  (1,1) circle(.25ex);
\draw[line width=.4pt]  (2,0) circle(.25ex);
\draw[line width=.4pt]  (2,2) circle(.25ex);
\draw[line width=.4pt]  (3,1) circle(.25ex);
\draw[line width=.4pt]  (4,0) circle(.25ex);
\draw[line width=.4pt]  (4,2) circle(.25ex);
\node[font=\scriptsize] at (-.4,2.03) {$P$};
\node[font=\scriptsize] at (2.45,.45) {$\mathcal R (P)$};
\end{scope}
\end{tikzpicture}
}
\caption{Three maximal slanted rectangles in the AR quiver of $\dq{A_3}$ indicating the nonzero Hom spaces from the objects marked by filled vertices to the objects in the rectangles}
\label{rectangle}
\end{figure}
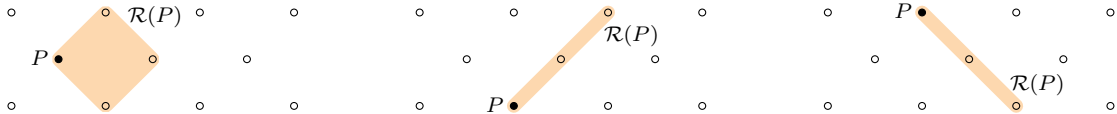

Given a $t$-structure $(\mathcal D^{\preceq 0}, \mathcal D^{\succeq 0})$ on a triangulated category $\mathcal D$ with heart $\mathcal H = \mathcal D^{\preceq 0} \cap \mathcal D^{\succeq 0}$, an object $P \in \mathcal D$ is called a projective of the heart if for all $M \in \mathcal H$ and all $k \neq 0$ one has $\operatorname{Hom}_{\mathcal D} (P, M [k]) = 0$. For type $A$ quivers the dimension of the Hom space between indecomposable objects can be read off the AR quiver \cite[\S 3.1.4]{schiffler} as $\operatorname{Hom} (P, M) \neq 0$ precisely when $M$ lies in the maximal slanted (possibly degenerate) rectangle $\mathcal R (P)$ whose left-most point is $P$ as illustrated for an $A_3$ quiver in Fig.~\ref{rectangle}.
A bounded $t$-structure is completely described by its heart,
which, in turn, is labelled by the projective objects of the heart. 

\subsection{Cluster category}
As mentioned earlier, the mesh \eq{mesh5:1} obtained from the planar variables upon
effecting the various identifications coincides with the cluster category of
type $A$ quivers. 
We now recall the notion of the cluster category for a quiver of type $A_n$ which is an orbit category of the derived category $\dq{A_n}$, 
obtained by quotienting with an automorphism of the derived
category. The example pertinent for us in the present article is the 
automorphism $F=\tau^{-1}\circ[1] \colon \dq{A_2}\rt\dq{A_2}$, called the 
cluster automorphism. Quotienting the derived category by $F$ yields a
cluster category. 
We shall discuss the notion with the example of  $\dq{A_2}$.

Including the AR translation $\tau$ indicating the existence of 
an extension of $1$ by $\tau (1) = 2$ furnishes  
the following diagrammatic picture of $\rep{A_2}$
\begin{equation}
\label{abelianAR}
\begin{tikzpicture}[baseline=-4pt]
\matrix (m) [matrix of math nodes, row sep={3em,between origins}, text height=1.4ex, column sep={3em,between origins}, text depth=.15ex, ampersand replacement=\&]
{
\& \nodetwo{1}{2} \& \\
2 \&\& 1 \rlap{.} \\
};
\draw[->] (m-2-1) -- (m-1-2);
\draw[->] (m-1-2) -- (m-2-3);
\path[->] (m-2-3) edge[dashed] node[above=-.3ex, font=\scriptsize] {$\tau$} (m-2-1);
\end{tikzpicture}
\end{equation}
Repeating this unit on both left and
right to include the shift functors (given diagrammatically 
by a glide reflection) yields the AR quiver \eq{dbA} corresponding
to the bounded derived category $\dq{A_2} = \db{A_2}$ whose objects 
are bounded complexes of representations in $\rep{A_2}$. 
Identifying the AR translation $\tau$ with the shift functor 
$[1]$ in $\dq{A_2}$ one obtains the cluster
category $\cq{A_2}$ \cite{bmrrt,ccs}. It is a triangulated category
\cite{keller} depicted as
\begin{equation}
\label{dbA2}
\begin{tikzpicture}[baseline=-4pt]
\matrix (m) [matrix of math nodes, minimum width=1.4em, row sep={3.6em,between origins}, text height=1.4ex, column sep={3.6em,between origins}, text depth=.15ex, ampersand replacement=\&]
{
\& \nodetwo{1}{2} \&\& \nodeone{2}[1] \&\& \nodeone{1}[1] \mathrlap{{}=F(\nodeone{2})} \&\&  \\
2 \&\& 1 \&\& \nodetwo{1}{2}[1] \&\& \nodeone{2}[2] \mathrlap{{}= F (\nodetwo{1}{2})} \& \phantom{\hspace{2.6em}} \\
};
\foreach \i/\j in {1/2,3/4,5/6}
\draw[->] (m-2-\i)--(m-1-\j);
%
\foreach \i/\j in {2/3,4/5,6/7}
\draw[->] (m-1-\i)--(m-2-\j);
%
\draw[->,dashed] (m-1-6) -- (m-1-4) node[above,midway,font=\scriptsize] {$\tau$};
\draw[->,dashed] (m-1-4) -- (m-1-2) node[above,midway,font=\scriptsize] {$\tau$};
\draw[->,dashed] (m-2-7) -- (m-2-5) node[above,midway,font=\scriptsize] {$\tau$};
\draw[->,dashed] (m-2-5) -- (m-2-3) node[above,midway,font=\scriptsize] {$\tau$};
\draw[->,dashed] (m-2-3) -- (m-2-1) node[above,midway,font=\scriptsize] {$\tau$};
\draw[opacity=.15,fill,line width=0] ([yshift=-1.6em]m-2-1.center) -- ++(7.2em,0) arc[start angle=-90, end angle=45, radius=1.7em] -- ++(-3.6em,3.6em) arc[start angle=45, end angle=135, radius=1.7em] -- ++(-3.6em,-3.6em) arc[start angle=135, end angle=270, radius=1.7em] -- cycle;
\node[font=\scriptsize,below=-.5ex of m-1-2] {$X_{14}$};
\node[font=\scriptsize,above=-.5ex of m-1-2] {$(0,2)$};
\node[font=\scriptsize,below=-.5ex of m-1-4] {$X_{25}$};
\node[font=\scriptsize,above=-.5ex of m-1-4] {$(1,2)$};
\node[font=\scriptsize,below=-.5ex of m-1-6] {$X_{13}$};
\node[font=\scriptsize,above=-.5ex of m-1-6] {$(2,2)$};
\node[font=\scriptsize,below=-.7ex of m-2-1] {$X_{13}$};
\node[font=\scriptsize,above=-.1ex of m-2-1] {$(0,1)$};
\node[font=\scriptsize,below=-.7ex of m-2-3] {$X_{24}$};
\node[font=\scriptsize,above=-.1ex of m-2-3] {$(1,1)$};
\node[font=\scriptsize,below=-.7ex of m-2-5] {$X_{35}$};
\node[font=\scriptsize,above=-.1ex of m-2-5] {$(2,1)$};
\node[font=\scriptsize,below=-.7ex of m-2-7] {$X_{14}$};
\node[font=\scriptsize,above=-.1ex of m-2-7] {$(3,1)$};
\end{tikzpicture}
\end{equation}
so that in the cluster category $\cq{A_2}$, $2$ is isomorphic to $F (2) = \nodeone{1}[1]$ and $\nodetwo{1}{2}$ is isomorphic to $F (\nodetwo{1}{2}) = \nodeone{2}[2]$.
Here the nodes are labelled by indecomposable representations 
in $\rep{A_2}$ and their shifts as in \eqref{shift}.
In \eqref{dbA2} we have also indicated the AR label\footnote{$(p,i)$ referring to the $i$th vertex in the $p$th copy of $Q$ in the translation quiver of $Q$}
above and the corresponding planar variable below each node. 
Objects in the cluster category satisfy mesh relations compatible with its
triangulated structure. 
The mesh relations written in terms of the AR labels are 
\cite{seyn,gph}
\begin{equation}
\label{mesh:r}
r_{(p,i)}= (p,i-1)\longrightarrow (p,i)\longrightarrow (p+1,i).
\end{equation} 
While the objects in the cluster category are orbits of complexes of quiver
representations under the cluster automorphism, we just need to work 
with the indecomposable objects, which we continue to denote simply by 
the indecomposable representations and their shifts. 

The cluster category captures the combinatorics of the intermediate $t$-structures or torsion pairs in terms of cluster tilting objects, which in the cluster category of a type $A$ quiver are precisely given by direct sums of projectives of the intermediate $t$-structures. Each term in the canonical form can thus be obtained from the cluster tilting objects, the indecomposable objects in the cluster category corresponding in some way to the planar variables.

We have remarked that the mesh \eq{mesh5:1} possesses the same structure of
vertices and arrows as the cluster category of ${A_2}$. However, the planar
variables are real numbers, satisfying the mesh relations obtained fron
\eq{mesh:X}, while the vertices in the cluster category are modules
satisfying the mesh relations \eq{mesh:r}. In order to relate them we 
interpret the planar variables as the modulus of central charges of these
objects. The planar variables then satisfy the $0$-deformed mesh relations
\cite{bdmty}.
The central charge is defined as a map from the Grothendieck group of the
derived category to the complex plane, 
$Z \colon K_0 (\dq{A_2})\rt\C$.
We label the central charge by the objects as well as the AR labels, 
for example, $Z_{\scalebox{.8}{$\nodetwo{1}{2}$}}=Z_{(0,2)}$
etc., and use the notations interchangeably. 
The mesh relations then give rise to relations among central charges as
\begin{equation}
\label{ccharg}
Z_{(p,i)}=Z_{(p,i-1)}+Z_{(p+1,i)}.
\end{equation} 
For the present instance, in $\dq{A_2}$, these are, with middle node 
$(0,2)$, $(1,1)$, $(1,2)$, $(2,1)$, $(2,2)$,
\begin{equation}
\label{mesh:a2}
\begin{gathered}
Z_{\scalebox{.8}{$\nodetwo{1}{2}$}}=Z_{\nodeone{2}}+Z_{\nodeone{1}},\\
Z_{\nodeone{1}}=Z_{\scalebox{.8}{$\nodetwo{1}{2}$}}+Z_{\nodeone{2}[1]},\\
Z_{\nodeone{2}[1]}=Z_{\nodeone{1}}+Z_{\scalebox{.8}{$\nodetwo{1}{2}$}[1]},\\
Z_{\scalebox{.8}{$\nodetwo{1}{2}$}[1]}=Z_{\nodeone{2}[1]}+Z_{\nodeone{1}[1]},\\
Z_{\nodeone{1}[1]}=Z_{\scalebox{.8}{$\nodetwo{1}{2}$}[1]}+Z_{\nodeone{2}[2]},
\end{gathered}
\end{equation} 
repectively.  We have five relations among seven central charges. 
Thus two of them are ``independent" and the rest
can be expressed as their linear combinations. 
The choice of the two independent ones correspond to a choice of intermediate $t$-structure, whose heart has in the case of $A_2$ two indecomposable projective objects. Once the independent ones
are chosen, the assignment of central charges is made according to
\begin{equation}
\label{Za1}
Z_{A[1]}=-Z_A
\end{equation}  
for any object $A$.

The reason for defining the charge on the Grothendieck group of the derived category is that the Grothendieck group of the cluster category with its standard triangulated structure is too small, as was brought to our attention by Yann Palu. This may be remedied by working with so-called extriangulated structures on the cluster category \cite{pppp}. For the present article we shall freely use the two equivalent perspectives of projectives hearts of intermediate $t$-structures on the one hand and direct summands of cluster tilting objects on the other hand, and content ourselves with defining the charge only from the projectives of hearts.
\subsection{Canonical form from hearts}
We have introduced the notions relevant for the categorification of the
canonical form. We now present the connection between its terms and the
projective objects  of the hearts of intermediate $t$-structures restricted
to the cluster category $\mathcal{C}_{A_n}$.

The standard heart in $\dq{A_2}$ is given by the Abelian category $\rep{A_2}$
\eqref{abelianAR} and the indecomposable projective objects of the 
heart are the representations $\nodeone{2}$ and $\nodetwo{1}{2}$ 
corresponding to the planar variables
$X_{13}$ and $X_{14}$, respectively, as indicated in \eq{dbA2}.
The corresponding  contribution to the 
canonical form \eqref{can:5} is $d \log X_{14} \wedge d \log X_{13}$. 

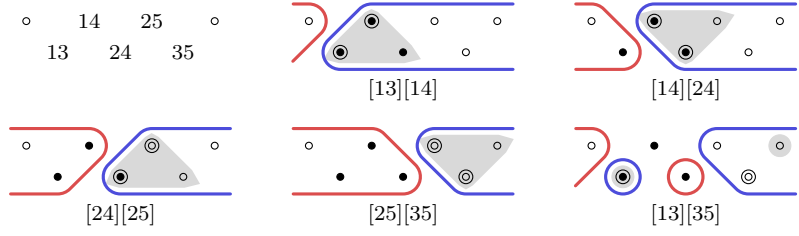
\begin{figure}
\begin{center}
\begin{tikzpicture}[x=1em,y=1em]
\draw[line width=.4pt] (-2,1) circle(.25ex);
\node[font=\scriptsize] at (-1,0) {$13$};
\node[font=\scriptsize] at  (0,1) {$14$};
\node[font=\scriptsize] at  (1,0) {$24$};
\node[font=\scriptsize] at  (2,1) {$25$};
\node[font=\scriptsize] at  (3,0) {$35$};
\draw[line width=.4pt]  (4,1) circle(.25ex);
\begin{scope}[xshift=9em]
\draw[line width=.7em,fill=black!15,black!15,rounded corners=.001em] (1,0) -- (-1,0) -- (0,1) -- cycle;
\draw[line width=.4pt] (-2,1) circle(.25ex);
\draw[line width=.4pt,fill=black] (-1,0) circle(.25ex);
\draw[line width=.4pt] (-1,0) circle(.5ex);
\draw[line width=.4pt,fill=black]  (0,1) circle(.25ex);
\draw[line width=.4pt]  (0,1) circle(.5ex);
\draw[line width=.4pt,fill=black]  (1,0) circle(.25ex);
\draw[line width=.4pt]  (2,1) circle(.25ex);
\draw[line width=.4pt]  (3,0) circle(.25ex);
\draw[line width=.4pt]  (4,1) circle(.25ex);
\draw[line width=.1em,red!80!black!70, line cap=round] (-2.5,1.55) -- (-2,1.55) arc[start angle=90, end angle=-45, radius=.55em] -- ++(-.9,-.9);
\draw[line width=.1em,blue!80!black!70,line cap=round] (4.5,1.55) -- (0,1.55) arc[start angle=90, end angle=135, radius=.55em] -- ++(-1,-1) arc[start angle=135, end angle=270, radius=.55em] -- ++(5.5,0);
\node[font=\scriptsize] at (1,-1.2) {$[13] [14]$};
\end{scope}
\begin{scope}[xshift=18em]
\draw[line width=.7em,fill=black!15,black!15,rounded corners=.001em] (1,0) -- (2,1) -- (0,1) -- cycle;
\draw[line width=.4pt] (-2,1) circle(.25ex);
\draw[line width=.4pt,fill=black] (-1,0) circle(.25ex);
\draw[line width=.4pt,fill=black]  (0,1) circle(.25ex);
\draw[line width=.4pt]  (0,1) circle(.5ex);
\draw[line width=.4pt,fill=black]  (1,0) circle(.25ex);
\draw[line width=.4pt]  (1,0) circle(.5ex);
\draw[line width=.4pt]  (2,1) circle(.25ex);
\draw[line width=.4pt]  (3,0) circle(.25ex);
\draw[line width=.4pt]  (4,1) circle(.25ex);
\draw[line width=.1em,red!80!black!70, line cap=round] (-2.5,1.55) -- (-2,1.55) arc[start angle=90, end angle=45, radius=.55em] -- ++(1,-1) arc[start angle=45, end angle=-90, radius=.55em] -- ++(-1.5,0);
\draw[line width=.1em,blue!80!black!70,line cap=round] (4.5,1.55) -- (0,1.55) arc[start angle=90, end angle=225, radius=.55em] -- ++(1,-1) arc[start angle=225, end angle=270, radius=.55em] -- ++(3.5,0);
\node[font=\scriptsize] at (1,-1.2) {$[14] [24]$};
\end{scope}
\begin{scope}[yshift=-4em]
\draw[line width=.7em,fill=black!15,black!15,rounded corners=.001em] (1,0) -- (2,1) -- (3,0) -- cycle;
\draw[line width=.4pt] (-2,1) circle(.25ex);
\draw[line width=.4pt,fill=black] (-1,0) circle(.25ex);
\draw[line width=.4pt,fill=black]  (0,1) circle(.25ex);
\draw[line width=.4pt,fill=black]  (1,0) circle(.25ex);
\draw[line width=.4pt]  (1,0) circle(.5ex);
\draw[line width=.4pt]  (2,1) circle(.25ex);
\draw[line width=.4pt]  (2,1) circle(.5ex);
\draw[line width=.4pt]  (3,0) circle(.25ex);
\draw[line width=.4pt]  (4,1) circle(.25ex);
\draw[line width=.1em,red!80!black!70, line cap=round] (-2.5,1.55) -- (0,1.55) arc[start angle=90, end angle=-45, radius=.55em] -- ++(-1,-1) arc[start angle=-45, end angle=-90, radius=.55em] -- ++(-1.5,0);
\draw[line width=.1em,blue!80!black!70,line cap=round] (4.5,1.55) -- (2,1.55) arc[start angle=90, end angle=135, radius=.55em] -- ++(-1,-1) arc[start angle=135, end angle=270, radius=.55em] -- ++(3.5,0);
\node[font=\scriptsize] at (1,-1.2) {$[24] [25]$};
\end{scope}
\begin{scope}[xshift=9em,yshift=-4em]
\draw[line width=.7em,fill=black!15,black!15,rounded corners=.001em] (3,0) -- (2,1) -- (4,1) -- cycle;
\draw[line width=.4pt] (-2,1) circle(.25ex);
\draw[line width=.4pt,fill=black] (-1,0) circle(.25ex);
\draw[line width=.4pt,fill=black]  (0,1) circle(.25ex);
\draw[line width=.4pt,fill=black]  (1,0) circle(.25ex);
\draw[line width=.4pt]  (2,1) circle(.25ex);
\draw[line width=.4pt]  (2,1) circle(.5ex);
\draw[line width=.4pt]  (3,0) circle(.25ex);
\draw[line width=.4pt]  (3,0) circle(.5ex);
\draw[line width=.4pt]  (4,1) circle(.25ex);
\draw[line width=.1em,red!80!black!70, line cap=round] (-2.5,1.55) -- (-0,1.55) arc[start angle=90, end angle=45, radius=.55em] -- ++(1,-1) arc[start angle=45, end angle=-90, radius=.55em] -- ++(-3.5,0);
\draw[line width=.1em,blue!80!black!70,line cap=round] (4.5,1.55) -- (2,1.55) arc[start angle=90, end angle=225, radius=.55em] -- ++(1,-1) arc[start angle=225, end angle=270, radius=.55em] -- ++(1.5,0);
\node[font=\scriptsize] at (1,-1.2) {$[25] [35]$};
\end{scope}
\begin{scope}[xshift=18em,yshift=-4em]
\draw[line width=0,fill=black!15,black!15] (-1,0) circle(.35em);
\draw[line width=0,fill=black!15,black!15] (4,1) circle(.35em);
\draw[line width=.4pt] (-2,1) circle(.25ex);
\draw[line width=.4pt,fill=black] (-1,0) circle(.25ex);
\draw[line width=.4pt] (-1,0) circle(.5ex);
\draw[line width=.4pt,fill=black]  (0,1) circle(.25ex);
\draw[line width=.4pt,fill=black]  (1,0) circle(.25ex);
\draw[line width=.4pt]  (2,1) circle(.25ex);
\draw[line width=.4pt]  (3,0) circle(.25ex);
\draw[line width=.4pt]  (3,0) circle(.5ex);
\draw[line width=.4pt]  (4,1) circle(.25ex);
\draw[line width=.1em,red!80!black!70, line cap=round] (-2.5,1.55) -- (-2,1.55) arc[start angle=90, end angle=-45, radius=.55em] -- ++(-.9,-.9);
\draw[line width=.1em,blue!80!black!70,line cap=round] (4.5,1.55) -- (2,1.55) arc[start angle=90, end angle=225, radius=.55em] -- ++(1,-1) arc[start angle=225, end angle=270, radius=.55em] -- ++(1.5,0);
\draw[line width=.1em,red!80!black!70,line cap=round] (1,0) circle(.55em);
\draw[line width=.1em,blue!80!black!70,line cap=round] (-1,0) circle(.55em);
\node[font=\scriptsize] at (1,-1.2) {$[13] [35]$};
\end{scope}
\end{tikzpicture}
\end{center}
\caption{The 5 intermediate $t$-structures of $\dq{A_2}$ obtained by tilting
with respect to a torsion pair from the standard heart (filled vertices), their hearts (shaded), projective
objects (circled vertices) and their contribution to the canonical form}
\label{hearts2}
\end{figure}
For the intermediate $t$-structures of $\dq{A_2}$ the projective objects of
their hearts are marked in Fig.~\ref{hearts2} as the circled vertices. From
the point of view of cluster categories, the direct sums of the images of the projectives of the hearts in the cluster category are precisely the cluster tilting objects of the cluster category $\cq{A_2}$. The corresponding planar variables contribute to the canonical form as follows.

\begin{itemize}
\item For the standard $t$-structure corresponding to the torsion pair $(\mathcal T, \mathcal F) = (\rep{A_2}, 0)$ (the first diagram in Fig.~\ref{torsionpairs2}), the heart is the standard heart $\rep{A_2} = \heart{\nodeone{2},\nodetwo{1}{2},\nodeone{1}}$ with (indecomposable) projectives $\nodeone{2}$ and $\nodetwo{1}{2}$ so that $\nodeone{2} \oplus \nodetwo{1}{2}$ is a cluster tilting object of $\cq{A_2}$, where we have written also $\nodeone{2}$ and $\nodetwo{1}{2}$ for their images in the cluster category. Their central charges are chosen to be the independent ones. All the others can be expressed in terms of them using \eq{mesh:a2}. We have,
for example,
\begin{equation}
Z_{\nodeone{1}}=
Z_{\scalebox{.8}{$\nodetwo{1}{2}$}}
-Z_{\nodeone{2}}.
\end{equation} 
Using \eq{Za1} all the equations \eq{mesh:a2} coincide with this, leaving
only two independent central charges.
Writing $Z_A = X_A \mathrm e^{i\pi\phi}$, for objects and choosing $X$'s according to
\eq{mesh:a2}, we can write this as a relation among the $X$'s,
\begin{equation}
X_{24}=X_{14}-X_{13},
\end{equation} 
where we have ignored the common phases of the central charges, tantamount to
choosing a line of stability. 
All the other $X$'s can be similarly 
expressed in terms of $X_{13}$ and $X_{14}$. Contribution to the canonical
form is $d\log X_{14}\wedge d\log X_{13}$.
\item The heart of the $t$-structure obtained by tilting with respect to the torsion pair $(\mathcal T, \mathcal F)$ with $\mathcal T = \langle \nodetwo{1}{2}, \nodeone{1} \rangle$ and $\mathcal F = \langle 2 \rangle$ is given by $\heart{\nodetwo{1}{2},\nodeone{1},\nodeone{2}[1]}$ whose projectives are $\nodeone{1}$ and $\nodetwo{1}{2}$ so that $\nodeone{1} \oplus \nodetwo{1}{2}$ is the corresponding cluster tilting object. (Note that here $2$ is no longer a projective of the heart of the corresponding intermediate $t$-structure, since $2 [1]$ is in the heart and $\operatorname{Hom}_{\dq{A_2}} (2, 2[1][-1]) = \operatorname{Hom} (2, 2) \simeq \C \neq 0$.) In this case the independent charges are $Z_{\nodeone{1}}$ and
$Z_{\scalebox{.8}{$\nodetwo{1}{2}$}}$. Correspondingly, the independent $X$'s are $X_{24}$ and $X_{14}$,
the rest being expressed in terms of these. Contribution to the canonical
form is $d\log X_{24}\wedge d\log X_{14}$.
\item The contribution to the canonical form is determined completely
analogously for the remaining three hearts, whose cluster tilting objects are
$\nodeone{1} \oplus \nodeone{2}[1]$, $\nodeone{2}[1] \oplus
\nodetwo{1}{2}[1]$ and $\nodeone{2} \oplus \nodetwo{1}{2}[1]$. Matching their
direct summands with the planar variables, their contributions to the
canonical form are given in Fig.~\ref{hearts2}, where we have omitted
$\wedge$ and written $[ij]$ for $d \log X_{ij}$. Note that the heart of the
$t$-structure of the last diagram in Fig.~\ref{hearts2} is not equivalent to
$\rep{A_2}$ and that one of the two indecomposable projectives of the heart
does not lie inside the heart. Indeed, this kind of phenomenon was 
part of the motivation for developing $\tau$-tilting theory \cite{air,ht} which generalizes the notion of tilting modules in a way that is compatible with mutations.
\end{itemize}
Collecting all these 
contributions we recover the canonical form \eq{can:5}. The signs of the terms are fixed, up to an overall factor, by
demanding invariance under simultaneous scaling of the planar 
variables \cite{abhy}. As mentioned in the introduction,
each term in the canonical form,
corresponding to a topologically different Feynman diagram, corresponds to an
intermediate $t$-structure, thereby justifying the use of the latter in the
scheme. 
\section{Example of $N=6$}
\label{sec:6}
We proceed similarly for the case of $N=6$ particles.
We identify the mesh diagram of the momenta
with a portion of the AR quiver of $\dq{A_3}$ corresponding to the cluster
category $\cq{A_3}$ which can be obtained by identifying the indecomposable objects $3 \simeq F (3) = 2 [1]$, $\nodetwo{2}{3} \simeq F (\nodetwo{2}{3}) = \nodetwo{1}{2}[1]$ and $\nodethree{1}{2}{3} \simeq F \bigl( \nodethree{1}{2}{3} \bigr) = \nodeone{3}[2]$ in the green block
\begin{equation}
\begin{tikzpicture}[baseline=-4pt]
\matrix (m) [matrix of math nodes, row sep={3em,between origins}, text
height=1.4ex, text centered, anchor=center,column sep={3em,between origins}, text depth=0.25ex, ampersand replacement=\&,
nodes in empty cells]
{
\nodeone{2}[-1] \&\& \nodeone{1}[-1] \&\& \nodethree{1}{2}{3} \&\& \nodeone{3}[1] \&\& \nodeone{2}[1] \&\& \nodeone{1}[1] \\
\& \nodetwo{1}{2}[-1] \&\& \nodetwo{2}{3} \&\& \nodetwo{1}{2} \&\& \nodetwo{2}{3}[1] \&\& \nodetwo{1}{2}[1] \& \\
\nodethree{1}{2}{3}[-1] \&\& \nodeone{3} \&\& \nodeone{2} \&\& \nodeone{1} \&\& \nodethree{1}{2}{3}[1] \&\& \nodeone{3}[2] \\
};
\foreach \i/\j in {1/2,3/4,5/6,7/8,9/10}
\draw[thin,->] (m-3-\i)--(m-2-\j);
\foreach \i/\j in {2/3,4/5,6/7,8/9,10/11}
\draw[thin,->] (m-2-\i)--(m-1-\j);
\foreach \i/\j in {1/2,3/4,5/6,7/8,9/10}
\draw[thin,->] (m-1-\i)--(m-2-\j);
\foreach \i/\j in {2/3,4/5,6/7,8/9,10/11}
\draw[thin,->] (m-2-\i)--(m-3-\j);
\draw[opacity=.15,fill,line width=0] ([yshift=-1.5em]m-3-3.center) -- ++(12em,0) arc[start angle=-90, end angle=45, radius=1.5em] -- ++(-6em,6em) arc[start angle=45, end angle=135, radius=1.5em] -- ++(-6em,-6em) arc[start angle=135, end angle=270, radius=1.5em] -- cycle;
\draw[line width=.15em,black!20!green!80!blue!70,line cap=round] ([yshift=-1.8em]m-3-3.center) -- ++(24em,0) arc[start angle=-90, end angle=45, radius=1.8em] -- ++(-6em,6em) arc[start angle=45, end angle=90, radius=1.8em] -- ++(-12em,0) arc[start angle=90, end angle=135, radius=1.8em] -- ++(-6em,-6em) arc[start angle=135, end angle=270, radius=1.8em] -- cycle;
\node[font=\scriptsize,below=-.7ex of m-3-3] {$X_{13}$};
\node[font=\scriptsize,below=-.3ex of m-2-4] {$X_{14}$};
\node[font=\scriptsize,below=-.3ex of m-1-5] {$X_{15}$};
\node[font=\scriptsize,below=-.7ex of m-3-5] {$X_{24}$};
\node[font=\scriptsize,below=-.3ex of m-2-6] {$X_{25}$};
\node[font=\scriptsize,below=-.3ex of m-1-7] {$X_{26}$};
\node[font=\scriptsize,below=-.7ex of m-3-7] {$X_{35}$};
\node[font=\scriptsize,below=-.3ex of m-2-8] {$X_{36}$};
\node[font=\scriptsize,below=-.3ex of m-1-9] {$X_{13}$};
\node[font=\scriptsize,below=-.3ex of m-3-9] {$X_{46}$};
\node[font=\scriptsize,below=-.3ex of m-2-10] {$X_{14}$};
\node[font=\scriptsize,below=-.7ex of m-3-11] {$X_{15}$};
\end{tikzpicture}
\end{equation} 
where the shaded part is the heart of the standard $t$-structure of
$\dq{A_3}$. There are now 14 torsion pairs, giving 14 intermediate
$t$-structures whose hearts have three projective objects each, which 
in turn correspond to the 14 cluster tilting objects. 
The canonical form has 14 terms 
\begin{multline}
\label{can:6}
\Omega_{(6)} = 
-[13][14][15]
+[15][25][24]
-[15][14][24]  
+[24][25][26]
+[26][25][35]\\
-[26][36][35]
+[26][36][46]
-[15][25][35]
+[13][14][46]
-[13][15][35]
\\
+[46][14][24]
-[13][36][46]
+[24][26][46]
-[35][36][13]
\end{multline}
where we denoted $[ij]=d\log X_{ij}$.
The hearts correspond to the fourteen vertices of an associahedron
\cite{kq,qiu12}.

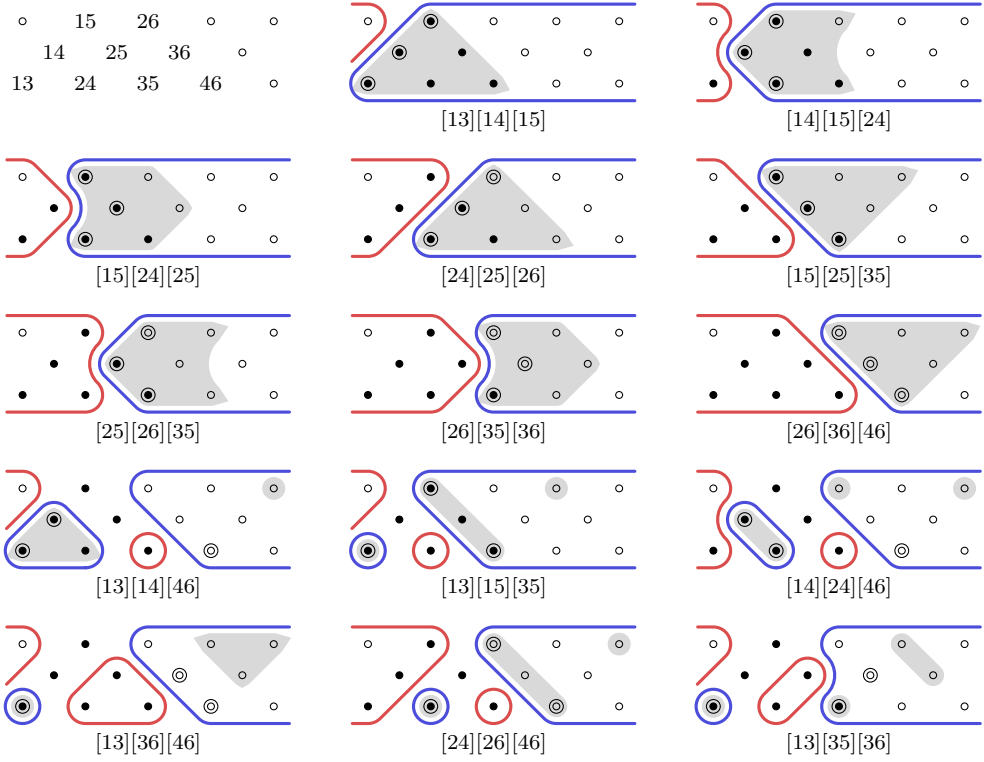
\begin{figure}[h]
\begin{center}
\begin{tikzpicture}[x=1em,y=1em]
\draw[line width=.4pt] (-2,2) circle(.25ex);
\node[font=\scriptsize] at (-2,0) {$13$};
\node[font=\scriptsize] at (-1,1) {$14$};
\node[font=\scriptsize] at  (0,2) {$15$};
\node[font=\scriptsize] at  (0,0) {$24$};
\node[font=\scriptsize] at  (1,1) {$25$};
\node[font=\scriptsize] at  (2,2) {$26$};
\node[font=\scriptsize] at  (2,0) {$35$};
\node[font=\scriptsize] at  (3,1) {$36$};
\node[font=\scriptsize] at  (4,0) {$46$};
\draw[line width=.4pt]  (4,2) circle(.25ex);
\draw[line width=.4pt]  (5,1) circle(.25ex);
\draw[line width=.4pt]  (6,0) circle(.25ex);
\draw[line width=.4pt]  (6,2) circle(.25ex);
\begin{scope}[xshift=11em]
\draw[line width=.7em,fill=black!15,black!15,rounded corners=.001em] (0,2) -- (2,0) -- (-2,0) -- cycle;
\draw[line width=.4pt] (-2,2) circle(.25ex);
\draw[line width=.4pt,fill=black] (-2,0) circle(.25ex);
\draw[line width=.4pt] (-2,0) circle(.5ex);
\draw[line width=.4pt,fill=black] (-1,1) circle(.25ex);
\draw[line width=.4pt] (-1,1) circle(.5ex);
\draw[line width=.4pt,fill=black]  (0,2) circle(.25ex);
\draw[line width=.4pt]  (0,2) circle(.5ex);
\draw[line width=.4pt,fill=black]  (0,0) circle(.25ex);
\draw[line width=.4pt,fill=black]  (1,1) circle(.25ex);
\draw[line width=.4pt,fill=black]  (2,0) circle(.25ex);
\draw[line width=.4pt]  (2,2) circle(.25ex);
\draw[line width=.4pt]  (3,1) circle(.25ex);
\draw[line width=.4pt]  (4,0) circle(.25ex);
\draw[line width=.4pt]  (4,2) circle(.25ex);
\draw[line width=.4pt]  (5,1) circle(.25ex);
\draw[line width=.4pt]  (6,0) circle(.25ex);
\draw[line width=.4pt]  (6,2) circle(.25ex);
\draw[line width=.1em,red!80!black!70, line cap=round] (-2.5,2.55) -- (-2,2.55) arc[start angle=90, end angle=-45, radius=.55em] -- ++(-.9,-.9);
\draw[line width=.1em,blue!80!black!70,line cap=round] (6.5,2.55) -- (0,2.55) arc[start angle=90, end angle=135, radius=.55em] -- ++(-2,-2) %
arc[start angle=135, end angle=270, radius=.55em] -- (6.5,-.55);
\node[font=\scriptsize] at (2,-1.2) {$[13] [14] [15]$};
\end{scope}
\begin{scope}[xshift=22em]
\draw[line width=.7em,fill=black!15,black!15,rounded corners=.001em] (2,0) -- (0,0) -- (-1,1) -- (0,2) -- (2,2) -- (1,1) -- cycle;
\draw[line width=.7em,black!15] (2,2) arc[start angle=135, end angle=225, radius=1.414em];
\draw[line width=.4pt] (-2,2) circle(.25ex);
\draw[line width=.4pt,fill=black] (-2,0) circle(.25ex);
\draw[line width=.4pt,fill=black] (-1,1) circle(.25ex);
\draw[line width=.4pt] (-1,1) circle(.5ex);
\draw[line width=.4pt,fill=black]  (0,0) circle(.25ex);
\draw[line width=.4pt]  (0,0) circle(.5ex);
\draw[line width=.4pt,fill=black]  (0,2) circle(.25ex);
\draw[line width=.4pt]  (0,2) circle(.5ex);
\draw[line width=.4pt,fill=black]  (1,1) circle(.25ex);
\draw[line width=.4pt,fill=black]  (2,0) circle(.25ex);
\draw[line width=.4pt]  (2,2) circle(.25ex);
\draw[line width=.4pt]  (3,1) circle(.25ex);
\draw[line width=.4pt]  (4,0) circle(.25ex);
\draw[line width=.4pt]  (4,2) circle(.25ex);
\draw[line width=.4pt]  (5,1) circle(.25ex);
\draw[line width=.4pt]  (6,0) circle(.25ex);
\draw[line width=.4pt]  (6,2) circle(.25ex);
\draw[line width=.1em,red!80!black!70, line cap=round] (-2.5,2.55) -- (-2,2.55) arc[start angle=90, end angle=-45, radius=.55em] %
arc[start angle=135, end angle=225, radius=.864em] arc[start angle=45, end angle=-90, radius=.55em] -- ++(-.5,0);
\draw[line width=.1em,blue!80!black!70,line cap=round] (6.5,2.55) -- (0,2.55) arc[start angle=90, end angle=135, radius=.55em] -- ++(-1,-1) arc[start angle=135, end angle=225, radius=.55em] -- ++(1,-1) arc[start angle=225, end angle=270, radius=.55em] -- (6.5,-.55);
\node[font=\scriptsize] at (2,-1.2) {$[14] [15] [24]$};
\end{scope}
\begin{scope}[yshift=-5em]
\draw[line width=.7em,fill=black!15,black!15,rounded corners=.001em] (2,0) -- (0,0) -- (1,1) -- (0,2) -- (2,2) -- (3,1) -- cycle;
\draw[line width=.7em,black!15] (0,2) arc[start angle=45, end angle=-45, radius=1.414em];
\draw[line width=.4pt] (-2,2) circle(.25ex);
\draw[line width=.4pt,fill=black] (-2,0) circle(.25ex);
\draw[line width=.4pt,fill=black] (-1,1) circle(.25ex);
\draw[line width=.4pt,fill=black]  (0,0) circle(.25ex);
\draw[line width=.4pt]  (0,0) circle(.5ex);
\draw[line width=.4pt,fill=black]  (0,2) circle(.25ex);
\draw[line width=.4pt]  (0,2) circle(.5ex);
\draw[line width=.4pt,fill=black]  (1,1) circle(.25ex);
\draw[line width=.4pt]  (1,1) circle(.5ex);
\draw[line width=.4pt,fill=black]  (2,0) circle(.25ex);
\draw[line width=.4pt]  (2,2) circle(.25ex);
\draw[line width=.4pt]  (3,1) circle(.25ex);
\draw[line width=.4pt]  (4,0) circle(.25ex);
\draw[line width=.4pt]  (4,2) circle(.25ex);
\draw[line width=.4pt]  (5,1) circle(.25ex);
\draw[line width=.4pt]  (6,0) circle(.25ex);
\draw[line width=.4pt]  (6,2) circle(.25ex);
\draw[line width=.1em,red!80!black!70, line cap=round] (-2.5,2.55) -- (-2,2.55) arc[start angle=90, end angle=45, radius=.55em] -- ++(1,-1) arc[start angle=45, end angle=-45, radius=.55em] -- ++(-1,-1) arc[start angle=-45, end angle=-90, radius=.55em] -- ++(-.5,0);
\draw[line width=.1em,blue!80!black!70,line cap=round] (6.5,2.55) -- (0,2.55) arc[start angle=90, end angle=225, radius=.55em] arc[start angle=45, end angle=-45, radius=.864em] arc[start angle=135, end angle=270, radius=.55em] -- ++(6.5,0);
\node[font=\scriptsize] at (2,-1.2) {$[15] [24] [25]$};
\end{scope}
\begin{scope}[xshift=11em,yshift=-5em]
\draw[line width=.7em,fill=black!15,black!15,rounded corners=.001em] (4,0) -- (0,0) -- (2,2) -- cycle;
\draw[line width=.4pt] (-2,2) circle(.25ex);
\draw[line width=.4pt,fill=black] (-2,0) circle(.25ex);
\draw[line width=.4pt,fill=black] (-1,1) circle(.25ex);
\draw[line width=.4pt,fill=black]  (0,0) circle(.25ex);
\draw[line width=.4pt]  (0,0) circle(.5ex);
\draw[line width=.4pt,fill=black]  (0,2) circle(.25ex);
\draw[line width=.4pt,fill=black]  (1,1) circle(.25ex);
\draw[line width=.4pt]  (1,1) circle(.5ex);
\draw[line width=.4pt,fill=black]  (2,0) circle(.25ex);
\draw[line width=.4pt]  (2,2) circle(.25ex);
\draw[line width=.4pt]  (2,2) circle(.5ex);
\draw[line width=.4pt]  (3,1) circle(.25ex);
\draw[line width=.4pt]  (4,0) circle(.25ex);
\draw[line width=.4pt]  (4,2) circle(.25ex);
\draw[line width=.4pt]  (5,1) circle(.25ex);
\draw[line width=.4pt]  (6,0) circle(.25ex);
\draw[line width=.4pt]  (6,2) circle(.25ex);
\draw[line width=.1em,red!80!black!70, line cap=round] (-2.5,2.55) -- (-0,2.55) arc[start angle=90, end angle=-45, radius=.55em] -- ++(-2,-2) arc[start angle=-45, end angle=-90, radius=.55em] -- ++(-.5,0);
\draw[line width=.1em,blue!80!black!70,line cap=round] (6.5,2.55) -- (2,2.55) arc[start angle=90, end angle=135, radius=.55em] -- ++(-2,-2) arc[start angle=135, end angle=270, radius=.55em] -- ++(6.5,0);
\node[font=\scriptsize] at (2,-1.2) {$[24] [25] [26]$};
\end{scope}
\begin{scope}[xshift=22em,yshift=-5em]
\draw[line width=.7em,fill=black!15,black!15,rounded corners=.001em] (4,2) -- (0,2) -- (2,0) -- cycle;
\draw[line width=.4pt] (-2,2) circle(.25ex);
\draw[line width=.4pt,fill=black] (-2,0) circle(.25ex);
\draw[line width=.4pt,fill=black] (-1,1) circle(.25ex);
\draw[line width=.4pt,fill=black]  (0,0) circle(.25ex);
\draw[line width=.4pt,fill=black]  (0,2) circle(.25ex);
\draw[line width=.4pt]  (0,2) circle(.5ex);
\draw[line width=.4pt,fill=black]  (1,1) circle(.25ex);
\draw[line width=.4pt]  (1,1) circle(.5ex);
\draw[line width=.4pt,fill=black]  (2,0) circle(.25ex);
\draw[line width=.4pt]  (2,0) circle(.5ex);
\draw[line width=.4pt]  (2,2) circle(.25ex);
\draw[line width=.4pt]  (3,1) circle(.25ex);
\draw[line width=.4pt]  (4,0) circle(.25ex);
\draw[line width=.4pt]  (4,2) circle(.25ex);
\draw[line width=.4pt]  (5,1) circle(.25ex);
\draw[line width=.4pt]  (6,0) circle(.25ex);
\draw[line width=.4pt]  (6,2) circle(.25ex);
\draw[line width=.1em,red!80!black!70, line cap=round] (-2.5,2.55) -- (-2,2.55) arc[start angle=90, end angle=45, radius=.55em] -- ++(2,-2) arc[start angle=45, end angle=-90, radius=.55em] -- ++(-2.5,0);
\draw[line width=.1em,blue!80!black!70,line cap=round] (6.5,2.55) -- (0,2.55) arc[start angle=90, end angle=225, radius=.55em] -- ++(2,-2) arc[start angle=225, end angle=270, radius=.55em] -- ++(4.5,0);
\node[font=\scriptsize] at (2,-1.2) {$[15] [25] [35]$};
\end{scope}
\begin{scope}[yshift=-10em]
\draw[line width=.7em,fill=black!15,black!15,rounded corners=.001em] (4,0) -- (2,0) -- (1,1) -- (2,2) -- (4,2) -- (3,1) -- cycle;
\draw[line width=.7em,black!15] (4,2) arc[start angle=135, end angle=225, radius=1.414em];
\draw[line width=.4pt] (-2,2) circle(.25ex);
\draw[line width=.4pt,fill=black] (-2,0) circle(.25ex);
\draw[line width=.4pt,fill=black] (-1,1) circle(.25ex);
\draw[line width=.4pt,fill=black]  (0,0) circle(.25ex);
\draw[line width=.4pt,fill=black]  (0,2) circle(.25ex);
\draw[line width=.4pt,fill=black]  (1,1) circle(.25ex);
\draw[line width=.4pt]  (1,1) circle(.5ex);
\draw[line width=.4pt,fill=black]  (2,0) circle(.25ex);
\draw[line width=.4pt]  (2,0) circle(.5ex);
\draw[line width=.4pt]  (2,2) circle(.25ex);
\draw[line width=.4pt]  (2,2) circle(.5ex);
\draw[line width=.4pt]  (3,1) circle(.25ex);
\draw[line width=.4pt]  (4,0) circle(.25ex);
\draw[line width=.4pt]  (4,2) circle(.25ex);
\draw[line width=.4pt]  (5,1) circle(.25ex);
\draw[line width=.4pt]  (6,0) circle(.25ex);
\draw[line width=.4pt]  (6,2) circle(.25ex);
\draw[line width=.1em,red!80!black!70, line cap=round] (-2.5,2.55) -- (0,2.55) arc[start angle=90, end angle=-45, radius=.55em] %
arc[start angle=135, end angle=225, radius=.864em] arc[start angle=45, end angle=-90, radius=.55em] -- ++(-2.5,0);
\draw[line width=.1em,blue!80!black!70,line cap=round] (6.5,2.55) -- (2,2.55) arc[start angle=90, end angle=135, radius=.55em] -- ++(-1,-1) arc[start angle=135, end angle=225, radius=.55em] -- ++(1,-1) arc[start angle=225, end angle=270, radius=.55em] -- (6.5,-.55);
\node[font=\scriptsize] at (2,-1.2) {$[25] [26] [35]$};
\end{scope}
\begin{scope}[xshift=11em,yshift=-10em]
\draw[line width=.7em,fill=black!15,black!15,rounded corners=.001em] (4,0) -- (2,0) -- (3,1) -- (2,2) -- (4,2) -- (5,1) -- cycle;
\draw[line width=.7em,black!15] (2,2) arc[start angle=45, end angle=-45, radius=1.414em];
\draw[line width=.4pt] (-2,2) circle(.25ex);
\draw[line width=.4pt,fill=black] (-2,0) circle(.25ex);
\draw[line width=.4pt,fill=black] (-1,1) circle(.25ex);
\draw[line width=.4pt,fill=black]  (0,0) circle(.25ex);
\draw[line width=.4pt,fill=black]  (0,2) circle(.25ex);
\draw[line width=.4pt,fill=black]  (1,1) circle(.25ex);
\draw[line width=.4pt,fill=black]  (2,0) circle(.25ex);
\draw[line width=.4pt]  (2,0) circle(.5ex);
\draw[line width=.4pt]  (2,2) circle(.25ex);
\draw[line width=.4pt]  (2,2) circle(.5ex);
\draw[line width=.4pt]  (3,1) circle(.25ex);
\draw[line width=.4pt]  (3,1) circle(.5ex);
\draw[line width=.4pt]  (4,0) circle(.25ex);
\draw[line width=.4pt]  (4,2) circle(.25ex);
\draw[line width=.4pt]  (5,1) circle(.25ex);
\draw[line width=.4pt]  (6,0) circle(.25ex);
\draw[line width=.4pt]  (6,2) circle(.25ex);
\draw[line width=.1em,red!80!black!70, line cap=round] (-2.5,2.55) -- (0,2.55) arc[start angle=90, end angle=45, radius=.55em] -- ++(1,-1) arc[start angle=45, end angle=-45, radius=.55em] -- ++(-1,-1) arc[start angle=-45, end angle=-90, radius=.55em] -- ++(-2.5,0);
\draw[line width=.1em,blue!80!black!70,line cap=round] (6.5,2.55) -- (2,2.55) arc[start angle=90, end angle=225, radius=.55em] arc[start angle=45, end angle=-45, radius=.864em] arc[start angle=135, end angle=270, radius=.55em] -- ++(4.5,0);
\node[font=\scriptsize] at (2,-1.2) {$[26] [35] [36]$};
\end{scope}
\begin{scope}[xshift=22em,yshift=-10em]
\draw[line width=.7em,fill=black!15,black!15,rounded corners=.001em] (6,2) -- (2,2) -- (4,0) -- cycle;
\draw[line width=.4pt] (-2,2) circle(.25ex);
\draw[line width=.4pt,fill=black] (-2,0) circle(.25ex);
\draw[line width=.4pt,fill=black] (-1,1) circle(.25ex);
\draw[line width=.4pt,fill=black]  (0,0) circle(.25ex);
\draw[line width=.4pt,fill=black]  (0,2) circle(.25ex);
\draw[line width=.4pt,fill=black]  (1,1) circle(.25ex);
\draw[line width=.4pt,fill=black]  (2,0) circle(.25ex);
\draw[line width=.4pt]  (2,2) circle(.25ex);
\draw[line width=.4pt]  (2,2) circle(.5ex);
\draw[line width=.4pt]  (3,1) circle(.25ex);
\draw[line width=.4pt]  (3,1) circle(.5ex);
\draw[line width=.4pt]  (4,0) circle(.25ex);
\draw[line width=.4pt]  (4,0) circle(.5ex);
\draw[line width=.4pt]  (4,2) circle(.25ex);
\draw[line width=.4pt]  (5,1) circle(.25ex);
\draw[line width=.4pt]  (6,0) circle(.25ex);
\draw[line width=.4pt]  (6,2) circle(.25ex);
\draw[line width=.1em,red!80!black!70, line cap=round] (-2.5,2.55) -- (0,2.55) arc[start angle=90, end angle=45, radius=.55em] -- ++(2,-2) arc[start angle=45, end angle=-90, radius=.55em] -- ++(-4.5,0);
\draw[line width=.1em,blue!80!black!70,line cap=round] (6.5,2.55) -- (2,2.55) arc[start angle=90, end angle=225, radius=.55em] -- ++(2,-2) arc[start angle=225, end angle=270, radius=.55em] -- ++(2.5,0);
\node[font=\scriptsize] at (2,-1.2) {$[26] [36] [46]$};
\end{scope}
\begin{scope}[yshift=-15em]
\draw[line width=.7em,fill=black!15,black!15,rounded corners=.001em] (0,0) -- (-2,0) -- (-1,1) -- cycle;
\draw[line width=0,fill=black!15,black!15] (6,2) circle(.35em);
\draw[line width=.4pt] (-2,2) circle(.25ex);
\draw[line width=.4pt,fill=black] (-2,0) circle(.25ex);
\draw[line width=.4pt] (-2,0) circle(.5ex);
\draw[line width=.4pt,fill=black] (-1,1) circle(.25ex);
\draw[line width=.4pt] (-1,1) circle(.5ex);
\draw[line width=.4pt,fill=black]  (0,0) circle(.25ex);
\draw[line width=.4pt,fill=black]  (0,2) circle(.25ex);
\draw[line width=.4pt,fill=black]  (1,1) circle(.25ex);
\draw[line width=.4pt,fill=black]  (2,0) circle(.25ex);
\draw[line width=.4pt]  (2,2) circle(.25ex);
\draw[line width=.4pt]  (3,1) circle(.25ex);
\draw[line width=.4pt]  (4,0) circle(.25ex);
\draw[line width=.4pt]  (4,0) circle(.5ex);
\draw[line width=.4pt]  (4,2) circle(.25ex);
\draw[line width=.4pt]  (5,1) circle(.25ex);
\draw[line width=.4pt]  (6,0) circle(.25ex);
\draw[line width=.4pt]  (6,2) circle(.25ex);
\draw[line width=.1em,red!80!black!70, line cap=round] (-2.5,2.55) -- (-2,2.55) arc[start angle=90, end angle=-45, radius=.55em] -- ++(-.9,-.9);
\draw[line width=.1em,red!80!black!70,line cap=round] (2,0) circle(.55em);
\draw[line width=.1em,blue!80!black!70,line cap=round] (6.5,2.55) -- (2,2.55) arc[start angle=90, end angle=225, radius=.55em] -- ++(2,-2) arc[start angle=225, end angle=270, radius=.55em] -- ++(2.5,0);
\draw[line width=.1em,blue!80!black!70,line cap=round] (0,-.55) -- ++(-2,0) arc[start angle=270, end angle=135, radius=.55em] -- ++(1,1) arc[start angle=135, end angle=45, radius=.55em] -- ++(1,-1) arc[start angle=45, end angle=-90, radius=.55em];
\node[font=\scriptsize] at (2,-1.2) {$[13] [14] [46]$};
\end{scope}
\begin{scope}[xshift=11em,yshift=-15em]
\draw[line width=0,fill=black!15,black!15] (-2,0) circle(.35em);
\draw[line width=0,fill=black!15,black!15]  (4,2) circle(.35em);
\draw[line width=.7em,fill=black!15,black!15,line cap=round] (0,2) -- (2,0);
\draw[line width=.4pt] (-2,2) circle(.25ex);
\draw[line width=.4pt,fill=black] (-2,0) circle(.25ex);
\draw[line width=.4pt] (-2,0) circle(.5ex);
\draw[line width=.4pt,fill=black] (-1,1) circle(.25ex);
\draw[line width=.4pt,fill=black]  (0,0) circle(.25ex);
\draw[line width=.4pt]  (0,2) circle(.5ex);
\draw[line width=.4pt,fill=black]  (0,2) circle(.25ex);
\draw[line width=.4pt,fill=black]  (1,1) circle(.25ex);
\draw[line width=.4pt,fill=black]  (2,0) circle(.25ex);
\draw[line width=.4pt]  (2,0) circle(.5ex);
\draw[line width=.4pt]  (2,2) circle(.25ex);
\draw[line width=.4pt]  (3,1) circle(.25ex);
\draw[line width=.4pt]  (4,0) circle(.25ex);
\draw[line width=.4pt]  (4,2) circle(.25ex);
\draw[line width=.4pt]  (5,1) circle(.25ex);
\draw[line width=.4pt]  (6,0) circle(.25ex);
\draw[line width=.4pt]  (6,2) circle(.25ex);
\draw[line width=.1em,red!80!black!70, line cap=round] (-2.5,2.55) -- (-2,2.55) arc[start angle=90, end angle=-45, radius=.55em] -- ++(-.9,-.9);
\draw[line width=.1em,red!80!black!70, line cap=round] (0,0) circle(.55em);
\draw[line width=.1em,blue!80!black!70,line cap=round] (6.5,2.55) -- (0,2.55) arc[start angle=90, end angle=225, radius=.55em] -- ++(2,-2) arc[start angle=225, end angle=270, radius=.55em] -- ++(4.5,0);
\draw[line width=.1em,blue!80!black!70,line cap=round] (-2,0) circle(.55em);
\node[font=\scriptsize] at (2,-1.2) {$[13] [15] [35]$};
\end{scope}
\begin{scope}[xshift=22em,yshift=-15em]
\draw[line width=.7em,fill=black!15,black!15,line cap=round] (0,0) -- (-1,1);
\draw[line width=0,fill=black!15,black!15] (2,2) circle(.35em);
\draw[line width=0,fill=black!15,black!15] (6,2) circle(.35em);
\draw[line width=.4pt] (-2,2) circle(.25ex);
\draw[line width=.4pt,fill=black] (-2,0) circle(.25ex);
\draw[line width=.4pt,fill=black] (-1,1) circle(.25ex);
\draw[line width=.4pt] (-1,1) circle(.5ex);
\draw[line width=.4pt,fill=black]  (0,0) circle(.25ex);
\draw[line width=.4pt]  (0,0) circle(.5ex);
\draw[line width=.4pt,fill=black]  (0,2) circle(.25ex);
\draw[line width=.4pt,fill=black]  (1,1) circle(.25ex);
\draw[line width=.4pt,fill=black]  (2,0) circle(.25ex);
\draw[line width=.4pt]  (2,2) circle(.25ex);
\draw[line width=.4pt]  (3,1) circle(.25ex);
\draw[line width=.4pt]  (4,0) circle(.25ex);
\draw[line width=.4pt]  (4,0) circle(.5ex);
\draw[line width=.4pt]  (4,2) circle(.25ex);
\draw[line width=.4pt]  (5,1) circle(.25ex);
\draw[line width=.4pt]  (6,0) circle(.25ex);
\draw[line width=.4pt]  (6,2) circle(.25ex);
\draw[line width=.1em,red!80!black!70, line cap=round] (-2.5,2.55) -- (-2,2.55) arc[start angle=90, end angle=-45, radius=.55em] %
arc[start angle=135, end angle=225, radius=.864em] arc[start angle=45, end angle=-90, radius=.55em] -- ++(-.5,0);
\draw[line width=.1em,red!80!black!70,line cap=round] (2,0) circle(.55em);
\draw[line width=.1em,blue!80!black!70,line cap=round] (6.5,2.55) -- (2,2.55) arc[start angle=90, end angle=225, radius=.55em] -- ++(2,-2) arc[start angle=225, end angle=270, radius=.55em] -- ++(2.5,0);
\draw[line width=.1em,blue!80!black!70,line cap=round] (0,-.55) arc[start angle=270, end angle=225, radius=.55em] -- ++(-1,1) arc[start angle=225, end angle=45, radius=.55em] -- ++(1,-1) arc[start angle=45, end angle=-90, radius=.55em];
\node[font=\scriptsize] at (2,-1.2) {$[14] [24] [46]$};
\end{scope}
\begin{scope}[yshift=-20em]
\draw[line width=0,fill=black!15,black!15] (-2,0) circle(.35em);
\draw[line width=.7em,fill=black!15,black!15,rounded corners=.001em] (4,2) -- (6,2) -- (5,1) -- cycle;
\draw[line width=.4pt] (-2,2) circle(.25ex);
\draw[line width=.4pt,fill=black] (-2,0) circle(.25ex);
\draw[line width=.4pt] (-2,0) circle(.5ex);
\draw[line width=.4pt,fill=black] (-1,1) circle(.25ex);
\draw[line width=.4pt,fill=black]  (0,0) circle(.25ex);
\draw[line width=.4pt,fill=black]  (0,2) circle(.25ex);
\draw[line width=.4pt,fill=black]  (1,1) circle(.25ex);
\draw[line width=.4pt,fill=black]  (2,0) circle(.25ex);
\draw[line width=.4pt]  (2,2) circle(.25ex);
\draw[line width=.4pt]  (3,1) circle(.25ex);
\draw[line width=.4pt]  (3,1) circle(.5ex);
\draw[line width=.4pt]  (4,0) circle(.25ex);
\draw[line width=.4pt]  (4,0) circle(.5ex);
\draw[line width=.4pt]  (4,2) circle(.25ex);
\draw[line width=.4pt]  (5,1) circle(.25ex);
\draw[line width=.4pt]  (6,0) circle(.25ex);
\draw[line width=.4pt]  (6,2) circle(.25ex);
\draw[line width=.1em,red!80!black!70, line cap=round] (-2.5,2.55) -- (-2,2.55) arc[start angle=90, end angle=-45, radius=.55em] -- ++(-.9,-.9);
\draw[line width=.1em,red!80!black!70,line cap=round] (2,-.55) -- ++(-2,0) arc[start angle=270, end angle=135, radius=.55em] -- ++(1,1) arc[start angle=135, end angle=45, radius=.55em] -- ++(1,-1) arc[start angle=45, end angle=-90, radius=.55em];
\draw[line width=.1em,blue!80!black!70,line cap=round] (6.5,2.55) -- (2,2.55) arc[start angle=90, end angle=225, radius=.55em] -- ++(2,-2) arc[start angle=225, end angle=270, radius=.55em] -- ++(2.5,0);
\draw[line width=.1em,blue!80!black!70, line cap=round] (-2,0) circle(.55em);
\node[font=\scriptsize] at (2,-1.2) {$[13] [36] [46]$};
\end{scope}
\begin{scope}[xshift=11em,yshift=-20em]
\draw[line width=0,fill=black!15,black!15]  (0,0) circle(.35em);
\draw[line width=0,fill=black!15,black!15]  (6,2) circle(.35em);
\draw[line width=.7em,fill=black!15,black!15,line cap=round] (2,2) -- (4,0);
\draw[line width=.4pt] (-2,2) circle(.25ex);
\draw[line width=.4pt,fill=black] (-2,0) circle(.25ex);
\draw[line width=.4pt,fill=black] (-1,1) circle(.25ex);
\draw[line width=.4pt,fill=black]  (0,0) circle(.25ex);
\draw[line width=.4pt]  (0,0) circle(.5ex);
\draw[line width=.4pt,fill=black]  (0,2) circle(.25ex);
\draw[line width=.4pt,fill=black]  (1,1) circle(.25ex);
\draw[line width=.4pt,fill=black]  (2,0) circle(.25ex);
\draw[line width=.4pt]  (2,2) circle(.25ex);
\draw[line width=.4pt]  (2,2) circle(.5ex);
\draw[line width=.4pt]  (3,1) circle(.25ex);
\draw[line width=.4pt]  (4,0) circle(.25ex);
\draw[line width=.4pt]  (4,0) circle(.5ex);
\draw[line width=.4pt]  (4,2) circle(.25ex);
\draw[line width=.4pt]  (5,1) circle(.25ex);
\draw[line width=.4pt]  (6,0) circle(.25ex);
\draw[line width=.4pt]  (6,2) circle(.25ex);
\draw[line width=.1em,red!80!black!70, line cap=round] (-2.5,2.55) -- (-0,2.55) arc[start angle=90, end angle=-45, radius=.55em] -- ++(-2,-2) arc[start angle=-45, end angle=-90, radius=.55em] -- ++(-.5,0);
\draw[line width=.1em,red!80!black!70, line cap=round] (2,0) circle(.55em);
\draw[line width=.1em,blue!80!black!70,line cap=round] (6.5,2.55) -- (2,2.55) arc[start angle=90, end angle=225, radius=.55em] -- ++(2,-2) arc[start angle=225, end angle=270, radius=.55em] -- ++(2.5,0);
\draw[line width=.1em,blue!80!black!70,line cap=round] (0,0) circle(.55em);
\node[font=\scriptsize] at (2,-1.2) {$[24] [26] [46]$};
\end{scope}
\begin{scope}[xshift=22em,yshift=-20em]
\draw[line width=0,fill=black!15,black!15] (-2,0) circle(.35em);
\draw[line width=0,fill=black!15,black!15]  (2,0) circle(.35em);
\draw[line width=.7em,fill=black!15,black!15,line cap=round] (4,2) -- (5,1);
\draw[line width=.4pt] (-2,2) circle(.25ex);
\draw[line width=.4pt,fill=black] (-2,0) circle(.25ex);
\draw[line width=.4pt] (-2,0) circle(.5ex);
\draw[line width=.4pt,fill=black] (-1,1) circle(.25ex);
\draw[line width=.4pt,fill=black]  (0,0) circle(.25ex);
\draw[line width=.4pt,fill=black]  (0,2) circle(.25ex);
\draw[line width=.4pt,fill=black]  (1,1) circle(.25ex);
\draw[line width=.4pt,fill=black]  (2,0) circle(.25ex);
\draw[line width=.4pt]  (2,0) circle(.5ex);
\draw[line width=.4pt]  (2,2) circle(.25ex);
\draw[line width=.4pt]  (3,1) circle(.25ex);
\draw[line width=.4pt]  (3,1) circle(.5ex);
\draw[line width=.4pt]  (4,0) circle(.25ex);
\draw[line width=.4pt]  (4,2) circle(.25ex);
\draw[line width=.4pt]  (5,1) circle(.25ex);
\draw[line width=.4pt]  (6,0) circle(.25ex);
\draw[line width=.4pt]  (6,2) circle(.25ex);
\draw[line width=.1em,red!80!black!70, line cap=round] (-2.5,2.55) -- (-2,2.55) arc[start angle=90, end angle=-45, radius=.55em] -- ++(-.9,-.9);
\draw[line width=.1em,red!80!black!70,line cap=round] (0,-.55) arc[start angle=270, end angle=135, radius=.55em] -- ++(1,1) arc[start angle=135, end angle=-45, radius=.55em] -- ++(-1,-1) arc[start angle=-45, end angle=-90, radius=.55em];
\draw[line width=.1em,blue!80!black!70,line cap=round] (6.5,2.55) -- (2,2.55) arc[start angle=90, end angle=225, radius=.55em] arc[start angle=45, end angle=-45, radius=.864em] arc[start angle=135, end angle=270, radius=.55em] -- ++(4.5,0);
\draw[line width=.1em,blue!80!black!70,line cap=round] (-2,0) circle(.55em);
\node[font=\scriptsize] at (2,-1.2) {$[13] [35] [36]$};
\end{scope}
\end{tikzpicture}
\end{center}
\caption{The 14 intermediate $t$-structures of $\dq{A_3}$ obtained by tilting from the standard heart (filled vertices), their hearts (shaded), projective objects (circled vertices) and their contribution to the canonical form}
\label{hearts3}
\end{figure}
\subsection{Cluster category, hearts and canonical form}
The torsion pairs for $\rep{A_3}$ and their corresponding $t$-structures for
$\dq{A_3}$ are illustrated in Fig.~\ref{hearts3}. (In each diagram the torsion class is the Abelian category generated by the black vertices in the blue part and the torsion-free class the one generated by the black vertices in the red part.) Let us remark that
the torsion pairs
of $\rep{A_3}$ can be ``pasted'' from the torsion pairs of $\rep{A_2}$ using
the following rule. First, a choice of a torsion pair on the three
lower left vertices and a torsion pair on the three lower right 
vertices has to be made
such that the overlap agrees. Since both $\mathcal T$ and $\mathcal F$ are closed under extensions and there can be no nonzero morphisms from objects in $\mathcal T$ to objects in $\mathcal F$. This determines in most cases uniquely whether the top vertex belongs to $\mathcal T$, to $\mathcal F$, or to neither. When there is a choice, both choices (i.e.\ including the top vertex in $\mathcal T$ or in $\mathcal F$) define torsion pairs. In this way one can obtain torsion pairs for $\rep{A_{n+1}}$ recursively from ``pasting'' torsion pairs of $\rep{A_n}$ and $\rep{A_2}$ (or equivalently from pasting $n$ torsion pairs of $\rep{A_2}$).

There are nine mesh relations ensuing from \eq{mesh:r} among the twelve
objects of the cluster category, namely,
\begin{equation}
\begin{tikzpicture}[baseline=-4pt]
\matrix (m) [matrix of math nodes, text height=1.8ex, text centered, row sep=.3em,
column sep=.9em, text depth=.6ex, ampersand replacement=\&] 
{
\nodeone{3} \& \nodetwo{2}{3} \& \nodeone{2}, \&[.8em]
\nodeone{2} \& \nodetwo{1}{2} \& \nodeone{1}, \&[.8em]
\nodeone{1} \& \nodetwo{2}{3}[1] \& \nodethree{1}{2}{3}[1], \\
\nodethree{1}{2}{3}[1] \& \nodetwo{1}{2}[1] \& \nodeone{3}[2], \&
\nodethree{1}{2}{3} \& \nodetwo{1}{2} \& \nodeone{3}[1], \&
\nodeone{3}[1] \& \nodetwo{2}{3}[1] \& \nodeone{2}[1], \\
\nodetwo{2}{3} \& \nodethree{1}{2}{3} \oplus \nodeone{2} \& \nodetwo{1}{2}, \&
\nodetwo{1}{2} \& \nodeone{3}[1] \oplus \nodeone{1} \& \nodetwo{2}{3}[1], \&
\nodetwo{2}{3}[1] \& \nodeone{2}[1] \oplus \nodethree{1}{2}{3}[1] \& \nodetwo{1}{2}[1]. \\
};
\draw[->,line width=.4pt] (m-1-1) -- (m-1-2);
\draw[->,line width=.4pt] (m-1-2) -- (m-1-3);
\draw[->,line width=.4pt] (m-1-4) -- (m-1-5);
\draw[->,line width=.4pt] (m-1-5) -- (m-1-6);
\draw[->,line width=.4pt] (m-1-7) -- (m-1-8);
\draw[->,line width=.4pt] (m-1-8) -- (m-1-9);
\draw[->,line width=.4pt] (m-2-1) -- (m-2-2);
\draw[->,line width=.4pt] (m-2-2) -- (m-2-3);
\draw[->,line width=.4pt] (m-2-4) -- (m-2-5);
\draw[->,line width=.4pt] (m-2-5) -- (m-2-6);
\draw[->,line width=.4pt] (m-2-7) -- (m-2-8);
\draw[->,line width=.4pt] (m-2-8) -- (m-2-9);
\draw[->,line width=.4pt] (m-3-1) -- (m-3-2);
\draw[->,line width=.4pt] (m-3-2) -- (m-3-3);
\draw[->,line width=.4pt] (m-3-4) -- (m-3-5);
\draw[->,line width=.4pt] (m-3-5) -- (m-3-6);
\draw[->,line width=.4pt] (m-3-7) -- (m-3-8);
\draw[->,line width=.4pt] (m-3-8) -- (m-3-9);
\end{tikzpicture}
\end{equation}

Accordingly, the twelve central charges are related by nine equations similar
to  \eq{mesh:a2}. The general rule of assignment of charges are
\begin{equation}
\label{aplusb} 
\begin{gathered}
Z_{A[1]}=-Z_A\\
Z_{A \oplus B} = Z_A + Z_B,
\end{gathered}
\end{equation} 
leaving three of them independent. As before, choosing the independent ones as 
the central charges
of the projectives of the hearts from Fig.~\ref{hearts3} their masses
furnish the terms of the canonical form \eq{can:6}. The signs are again fixed
by the scheme described in section~\ref{sec:N5}, which guarantees its
invariance under the scaling of the planar variables. 
\section{Conclusion}
\label{sec:concl}
In this note we have considered the canonical form appearing in the
computation of planar tree level Feynman diagrams of a cubic scalar 
field theory. The canonical form is a  means to encode
the contribution of these Feynman diagrams to the scattering amplitude.
Its relation to various mathematical structures have been studied earlier. In
here we interpret the terms of the canonical form as arising from the cluster tilting objects of the cluster categories of quivers of type $A$ which correspond to projectives of hearts of intermediate $t$-structures. 
This approach is categorical and makes no allusion to associahedrons 
and triangulations of polygons, although there is a precise general relation, the details of which we presented in two generic examples.
As mentioned in the introduction, each $N$-particle planar diagram for any
$N$ in a cubic theory corresponds to a triangulation of an $N$-gon by
non-intersecting diagonals. The category of diagonals of an $N$-gon is
identified with the cluster category of an $A_{N-3}$ quiver \cite{ccs,baur}. 
The associahendron is then obtained as the exchange graph of the hearts of
the intermediate $t$-structures of the cluster category \cite{kq,qiu12}.
The objects in the cluster category, which are representations of the
$A_{N-3}$ quiver, are then mapped to real numbers using the central charge,
as described in the text. This furnishes the rationale of identifying the
planar variables $X_{ij}$ arising in the kinematics of scattering with the
mass of the central charges via mesh relations. The terms in the canonical
form is then shown to be expressed in terms of the projectives of the
hearts of intermediate $t$-structures,
or equivalently of the direct summands of
cluster tilting objects in the cluster category. 
The present treatment also generalizes to quadratic and
higher order scalar field theories \cite{bopr} in terms of higher cluster categories.  

Each Feynman diagram contributes a term in the canonical form or scattering
amplitude. As demonstrated here, each of these corresponds to an intermediate 
$t$-structure and hence to a specific stability
regime.  We have illustrated this in two examples. The first one
is for $N=5$ particles, which is simpler, if somewhat restricted. The second
example of $N=6$ particles corresponding to the cluster category of $A_3$
quivers is generic. These considerations can be generalized to arbitrary
number of particles. The categorification is
expected to help the organization of the canonical form, especially in their
evaluation using computer programs. 

\section*{Acknowledgement}
We warmly thank Hipolito Treffinger for very helpful discussions and Yann Palu for several helpful comments and for bringing the reference \cite{pppp} to our attention.

\end{document}